\documentclass[%
    aps,
    prd,
    letterpaper,
    superscriptaddress,
    preprintnumbers,
    floatfix,
    twocolumn,
    nofootinbib,
    showpacs,
    hyphens
]{revtex4-1}  

\usepackage[normalem]{ulem}
\usepackage{amsmath,amsfonts,amssymb,amsthm}
\usepackage{subfigure}
\usepackage{graphicx}
\usepackage{dcolumn}
\usepackage{bm}
\usepackage{hyperref}
\usepackage[capitalise]{cleveref}
\usepackage{xcolor}
\newcommand{\ket}[1]{\left| #1 \right>} 
\newcommand{\bra}[1]{\left< #1 \right|} 
\newcommand{\braket}[2]{\left< #1 \vphantom{#2} \right|
 \left. #2 \vphantom{#1} \right>} 
\newcommand{\mbraket}[3]{\left< #1 \vphantom{#2#3} \right|
 #2 \left| #3 \vphantom{#1#2} \right>} 

\usepackage{slashed}

\usepackage{duckuments}

\newcommand{\bs}{\boldsymbol}

\newcommand{\Ham}{H}
\newcommand{\Vop}{V}
\newcommand{\Top}{T}

\begin{document}

\preprint{FERMILAB-PUB-25-0423-T}

\title{Baryon-baryon, meson-meson, and meson-baryon interactions in nonrelativistic QCD}
\author{Beno\^{i}t Assi}
\affiliation{Fermi National Accelerator Laboratory, Batavia, IL, 60510}
\affiliation{Department of Physics, University of Cincinnati, Cincinnati, Ohio 45221, USA}
\author{Anthony~V.~Grebe}
\affiliation{Fermi National Accelerator Laboratory, Batavia, IL, 60510}
\author{Michael~L.~Wagman}
\affiliation{Fermi National Accelerator Laboratory, Batavia, IL, 60510}

\date{\today}

\begin{abstract}
Van der Waals potentials describing interactions between color-singlet mesons and/or baryons vanish at leading order in potential nonrelativistic quantum chromodynamics (pNRQCD).
This result and constraints from Gauss's law are used to prove that weakly-coupled pNRQCD van der Waals potentials in generic non-Abelian gauge theories with only heavy quarks are too weak to form bound states whose color state is a product of color-singlets.
Quantum Monte Carlo calculations of four, five, and six quarks with equal masses provide numerical evidence that exotic color configurations are higher energy than products of color-singlet hadrons, suggesting that equal-mass fully-heavy tetraquark, pentaquark, and hexaquark bound states do not exist at next-to-leading order in pNRQCD and at all orders in QCD-like theories in which all quark masses are asymptotically large.
Mechanisms for generating hadron-hadron bound states are identified, which necessarily involve large quark-mass hierarchies, relativistic effects arising from the presence of sufficiently light quarks, or nonperturbative effects outside the scope of weakly-coupled pNRQCD. 
\end{abstract}

\maketitle

\section{Introduction}
\label{sec:intro}

Understanding the quark-mass dependence of hadron structure and interactions in quantum chromodynamics (QCD) is a long-standing theoretical challenge.
It has applications in lattice QCD, where numerical calculations are performed at a variety of masses to interpolate or extrapolate to the physical point, as well as to constrain new physics, including strongly coupled dark sectors.
More broadly, understanding the degree to which our universe is finely tuned requires understanding what range of Standard Model parameters can support the binding of hadrons into nuclei, and therefore the emergence of the Periodic Table and chemistry.

Meson-meson and meson-baryon interactions in the $m_Q \ll \Lambda_{\rm QCD}$ limit have been studied extensively with chiral perturbation theory~\cite{Bernard:1995dp,Scherer:2002tk,Bernard:2006gx}.
The meson-meson and meson-baryon scattering lengths are proportional to $\sqrt{ m_Q }$ in this regime and vanish in the chiral limit~\cite{Weinberg:1966kf,Tomozawa:1966jm}.
Baryon-baryon interactions are more challenging to study in chiral effective field theories.
The deuteron bound state present in nature may exist in the chiral limit with a somewhat larger binding energy~\cite{Beane:2002vs,Epelbaum:2002gb,Beane:2002xf,Berengut:2013nh}, although systematic uncertainties arising from the pion-mass dependence of short-range interactions in the chiral Lagrangian are challenging to quantify.

Here, we consider the opposite limit of $m_Q \gg \Lambda_{\rm QCD}$.
Even though this limit is obviously not applicable to the light and strange quarks present in nature, understanding this limit may shed light on lattice QCD calculations of two-baryon systems with $m_Q \sim \Lambda_{\rm QCD}$~\cite{Fukugita:1994na,Fukugita:1994ve,Beane:2006mx,Beane:2006gf,Beane:2009py,Yamazaki:2009ua,Inoue:2010hs,Inoue:2010es,NPLQCD:2010ocs,Inoue:2011ai,NPLQCD:2011naw,NPLQCD:2012mex,Yamazaki:2012hi,NPLQCD:2013bqy,Ishii:2013cta,Yamazaki:2015asa,Berkowitz:2015eaa,Wagman:2017tmp,Gongyo:2017fjb,Francis:2018qch,Horz:2020zvv,Green:2021qol,Amarasinghe:2021lqa,Detmold:2024iwz,BaSc:2025yhy} and heavy quarks~\cite{Junnarkar:2019equ,Lyu:2021qsh,Mathur:2022ovu,junnarkar:2022yak,Dhindsa:2025gae}, phenomenologically viable theories of dark matter involving dark sector analogs of heavy quarks~\cite{Mitridate:2017oky,Cline:2016nab,DeGrand:2019vbx,Cline:2021itd,Asadi:2021yml,Asadi:2021pwo,Asadi:2021bxp, Asadi:2024tpu, Asadi:2024bbq,Bodas:2024idn,Gouttenoire:2023roe}, and constraints on new physics involving big-bang nucleosynthesis~\cite{Dixit:1987at,Yoo:2002vw,Dmitriev:2003qq,Muller:2004gu,Chamoun:2005xr,Coc:2006sx,Landau:2008re,Bedaque:2010hr,Burns:2024ods,Meyer:2024auq,Meissner:2025dnx}.
For bound states comprised of heavy quarks, QCD dynamics are more straightforward due to the large hierarchy between the quark mass $m_Q$ and the nonperturbative scale $\Lambda_{\rm QCD}$ and can be studied using effective field theory (EFT)~\cite{Bodwin:1994jh,Caswell:1985ui,Pineda:1997bj,Pineda:1998kj,Brambilla:1999xf,Brambilla:2004jw}.
Quark velocities are small in such systems, $v \ll 1$, leading to a clear hierarchy of scales: $m_Q\gg p\sim m_Qv\gg E\sim m_Q v^2$~\cite{Caswell:1985ui}.
Integrating out the hard scale $m_Q$ leads to nonrelativistic QCD (NRQCD)~\cite{Bodwin:1994jh,Caswell:1985ui,Pineda:1997bj,Pineda:1998kj}, while further integrating out the soft scale $p_Q\sim m_Q v$ leads to potential NRQCD (pNRQCD)~\cite{Brambilla:1999xf}.
This soft scale sets the typical bound state size, which is analogous to the Bohr radius of the hydrogen atom. In the weak coupling regime of pNRQCD~\cite{Pineda:2011dg}, dynamics at the soft scale are incorporated by solving the time-independent Schr{\"o}dinger equation with a potential that incorporates all pNRQCD effects that are enhanced for small $p / m_Q$ and must be treated nonperturbatively.
Studies of heavy quarkonium have shown that pNRQCD can accurately describe properties of fully heavy quark and antiquark bound states, including masses and decay widths~\cite{Brambilla:1999xj,Kniehl:2002br,Brambilla:1999xf,Pineda:2011dg,Pineda:1997hz,Brambilla:2009bi, Brambilla:2005zw, Brambilla:2000db,Brambilla:2012be}.

The potentials needed to describe more complex systems, such as baryons, have been studied more recently~\cite{Savage:1990di,Brambilla:2005yk,Brambilla:2009cd,Brambilla:2013vx,Assi:2023cfo}. Recently, Ref.~\cite{Assi:2023cfo} extended the potential nonrelativistic QCD (pNRQCD) framework to generic multihadron systems using the heavy quark degrees of freedom as in velocity NRQCD (vNRQCD)~\cite{Luke:1999kz,Manohar:2000kr,Hoang:2002ae}.
Variational methods have subsequently been used to bound fully-heavy baryon masses~\cite{Jia:2006gw,Llanes-Estrada:2011gwu,Assi:2023cfo}, and Green's function Monte Carlo (GFMC) methods used to solve quantum many-body problems in nuclear and condensed matter physics~\cite{Carlson:2014vla,Yan_2017,Gandolfi:2020pbj} were further used to compute baryon masses in pNRQCD in Ref.~\cite{Assi:2023cfo}.
Variational and GFMC methods were applied to four-quark systems in pNRQCD at next-to-leading order (NLO) in Ref.~\cite{Assi:2023dlu}.
Tetraquark bound states were observed to exist if and only if the heavy quark/antiquark mass ratio is larger than a critical value, which is consistent with a previous variational study of one-gluon-exchange potentials equivalent to pNRQCD at leading order (LO)~\cite{Czarnecki:2017vco}.

In this work, we prove the non-existence of fully-heavy hadron-hadron bound states whose color states are products of color-singlet hadrons in the weakly coupled pNRQCD regime using all-orders arguments that constrain the form of non-Abelian ``van der Waals'' potentials.
We search for pentaquark and hexaquark bound states with exotic color structures in pNRQCD using GFMC calculations involving a complete basis of five- and six-quark color states at LO and NLO.
We further study the form of van der Waals potentials that arise between color-singlet hadrons in pNRQCD at NNLO.

The remainder of this paper is organized as follows.
\cref{sec:vdwTheory} presents all-orders arguments about the structure of van der Waals potentials and non-existence of color-singlet product bound states that are valid for generic non-Abelian gauge theories.
\cref{sec:NLO} discusses pNRQCD potentials for multihadron states and presents GFMC results for five- and six-quark bound states at LO and NLO.
\cref{sec:NNLO} explores pNRQCD van der Waals potentials that arise at NNLO and verifies their multipole expansion is consistent with our all-orders results.
\cref{sec:conclusion} discusses some implications of our results.


\section{Color-singlet product states in non-Abelian gauge theories}\label{sec:vdwTheory}

This section provides a proof of the following result: $SU(N)$, $SO(N)$, or $Sp(2N)$ gauge theories do not have bound states whose color structure is a product of two color-singlet hadrons when all colored fermion and scalar masses are asymptotically large. The result is valid at all orders of the weakly-coupled pNRQCD expansion. Note that the arguments in this section strictly apply only in the asymptotic heavy-mass limit (parametrically large masses for all colored fermions and scalars) and both the existence of light quark flavors and range of convergence of weakly coupled pNRQCD add complications for describing finite-mass heavy flavors (e.g., $b$ quarks) in real-world QCD.

\subsection{Van der Waals potentials vanish at LO}\label{sec:LO}

At LO in pNRQCD, the potentials between colored particles are given by contractions of one-gluon-exchange tree diagrams with the color tensors describing the external state.
For the non-Abelian gauge groups $SU(N)$, $SO(N)$, or $Sp(2N)$, the color structure of these ``one-gluon-exchange'' diagrams is given by contractions of $T^a \otimes T^a$ and its transposes. 
Here, the $T^a$ are Lie algebra generators satisfying the tracelessness condition $\text{Tr}[T^a] = 0$ that ensures that group elements have unit determinant and a conventional normalization condition chosen here as $\text{Tr}[T^a T^b] = \frac{1}{2} \delta^{ab}$.
Unit-normalized tensors for color-singlet meson and baryon states are given by $\delta_{ij} /\sqrt{N}$ and $\varepsilon_{ijk\ldots} / \sqrt{N!}$, respectively.
The color-singlet $Q\overline{Q}$ potential is given by $-\alpha / r$, where $\alpha = g^2/(4\pi)$ is the gauge coupling, times the contraction of $T^a \otimes [T^a]^T$ with color-singlet meson tensors,
\begin{equation}
  \frac{\delta_{ij} T^a_{ii'} T^a_{j'j} \delta_{i'j'}}{N} = \frac{\text{Tr}[T^a T^a]}{N} = \frac{d_{\rm Adj}}{2 N},
\end{equation}
where $d_{\rm Adj}$ is the dimension of the adjoint representation, which, e.g., for the $SU(N)$ case relevant to QCD is $N^2 - 1$.
The color-antisymmetric $QQ$ potential is given similarly by $-\alpha / r$ times the contraction of $T^a \otimes T^a$ with color-singlet baryon tensors,
\begin{equation}
  -\frac{\varepsilon_{ijk\ldots} T^a_{ii'} T^a_{jj'} \varepsilon_{i'j'k'\ldots}}{N!} = \frac{\text{Tr}[T^a T^a]}{N(N-1)} = \frac{d_{\rm Adj}}{2 N(N-1)}.
\end{equation}

The potentials for meson-meson product states involve contractions of $T^a \otimes T^a$ and its transposes with a color tensor proportional to $\delta_{ij} \delta_{kl}$.
Contractions involving the $Q\overline{Q}$ pair within each meson lead to the same color-singlet potential above.
However, the contractions between a $Q$ in one meson and $\overline{Q}$ in another meson vanish,
  \begin{equation}\label{eq:tree_vanishing_MM}
    \begin{split}
      & \delta_{ij} \delta_{kl} T^a_{jj'} T^a_{k'k} \delta_{ij'} \delta_{k'l} \\
      &= (\delta_{ij}\delta_{ij'}) (\delta_{kl} \delta_{k'l}) T^a_{jj'}T^a_{k'k}   \\
      &\propto \delta_{jj'}\delta_{kk'} T^a_{jj'}T^a_{k'k} = \text{Tr}[T^a]\text{Tr}[T^a] \\
      &= 0,
    \end{split}
  \end{equation}
due to the tracelessness of the generators.
Contractions involving a $Q$ in one meson and $Q$ in the other meson vanish analogously,
  \begin{equation}\label{eq:tree_vanishing_MM_2}
    \begin{split}
      & \delta_{ij} \delta_{kl} T^a_{jj'} T^a_{ll'} \delta_{ij'} \delta_{kl'} \\
      &= (\delta_{ij}\delta_{ij'}) (\delta_{kl} \delta_{kl'}) T^a_{jj'}T^a_{ll'}   \\
      &\propto \delta_{jj'}\delta_{ll'} T^a_{jj'}T^a_{ll'} = \text{Tr}[T^a]\text{Tr}[T^a] \\
      &= 0,
    \end{split}
  \end{equation}
as do contractions between the $\overline{Q}$ in each meson.
This means that the matrix element of the potential in a meson state $\ket{ \overline{Q}(\bm{x}_1) Q(\bm{x}_2) \overline{Q}(\bm{x}_3) Q(\bm{x}_4) }$ is given simply by the sum of the two intra-meson potentials,
\begin{equation}
  \mbraket{ \overline{Q} Q \overline{Q} Q }{V^{\rm LO}}{ \overline{Q} Q \overline{Q} Q } = -\frac{d_{\rm Adj}}{2N}\left( \frac{\alpha}{|\bm{r}_{12}|} + \frac{\alpha}{|\bm{r}_{34}|} \right).
\end{equation}
This can be contrasted with the case of positronium-positronium product states in QED, 
\begin{equation}
  \begin{split}
    \mbraket{ e^+ e^- e^+ e^- }{V^{\rm QED}}{ e^+ e^- e^+ e^- } =&  - \frac{\alpha_{\rm em}}{|\bm{r}_{12}|} - \frac{\alpha_{\rm em}}{|\bm{r}_{34}|}  \\
    &- \frac{\alpha_{\rm em}}{|\bm{r}_{14}|} - \frac{\alpha_{\rm em}}{|\bm{r}_{23}|}  \\
    &+ \frac{\alpha_{\rm em}}{|\bm{r}_{13}|} + \frac{\alpha_{\rm em}}{|\bm{r}_{24}|} .
  \end{split}
\end{equation}
The second two lines in the QED case are van der Waals potentials, which are present between charged constituents of neutral atoms, positronia, and other composite particles in QED.
Tracelessness of the non-Abelian generators leads to the vanishing of analogous van der Waals potentials at LO for meson-meson product states in $SU(N)$, $SO(N)$, and $Sp(2N)$ gauge theories.

An identical cancellation arises for van der Waals potentials between color-singlet baryons, because
\begin{equation}\label{eq:tree_vanishing_BB}
    \begin{split}
      & \varepsilon_{ijk...} \varepsilon_{lmn...} T^a_{ii'}T^a_{ll'} \varepsilon_{i'jk...} \varepsilon_{l'mn...} \\
      &= (\varepsilon_{ijk...}\varepsilon_{i'jk...} ) (\varepsilon_{lmn...}\varepsilon_{l'mn...}) T^a_{ii'}T^a_{ll'}   \\
      &\propto \delta_{ii'}\delta_{ll'} T^a_{ii'}T^a_{ll'} = \text{Tr}[T^a]\text{Tr}[T^a]\\
      &= 0.
    \end{split}
  \end{equation}
For meson-baryon systems, the analogous result is
  \begin{equation}\label{eq:tree_vanishing_MB}
    \begin{split}
    & \delta_{ij} \varepsilon_{klm...} T^a_{jj'} T^a_{kk'} \delta_{ij'} \varepsilon_{k'lm...} \\
      &= (\delta_{ij}\delta_{ij'}) (\varepsilon_{klm...}\varepsilon_{k'lm...}) T^a_{jj'}T^a_{kk'}   \\
      &\propto \delta_{jj'}\delta_{kk'} T^a_{jj'}T^a_{kk'} = \text{Tr}[T^a]\text{Tr}[T^a] \\
      &= 0.
    \end{split}
  \end{equation}
Spin- and flavor-dependent potentials are suppressed by powers of $1/m_Q$ as well as an additional power of $\alpha$.
In particular, the same LO potential applies to nonrelativistic scalars as it does to fermions.
This argument therefore demonstrates the vanishing of $O(\alpha)$ van der Waals potentials for any dark sector ``hadrons'' comprised of fermions or scalars charged under an arbitrary gauge group with traceless generators.

This cancellation implies that the product of any two single-hadron energy eigenstates is itself an energy eigenstate at LO.
To see this, note that the potential terms involving quarks within each hadron precisely cancel the kinetic energy contributions of the same quarks by the fact that each hadron wavefunction solves the Schr{\"o}dinger equation.
Because there are no additional potential terms involving quarks in different hadrons, the two-hadron product wavefunction is also a solution to the Schr{\"o}dinger equation.
The mass of this two-hadron eigenstate is equal to twice the mass of the individual hadrons; however, this does not rule out the possibility that a lower-energy eigenstate describing a two-hadron bound state also exists, as discussed below.

Previous works have studied van der Waals potentials in QED and QCD using effective theory frameworks, including pNRQCD~\cite{Brambilla:2017ffe}, 
and in particular there is a long history of work on understanding quarkonium-nucleus bound states that could arise from QCD van der Waals forces~\cite{Brodsky:1989jd,Luke:1992tm,Brodsky:1997yr,Brodsky:1997gh}.
It is well-known that net one-gluon-exchange potentials vanish between color-singlet hadrons
~\cite{Peskin:1979va,Bhanot:1979vb} and chromopolarizability effects associated with two-gluon-exchange have been studied extensively for this reason~\cite{Brambilla:2015rqa,Brambilla:2019esw,Dong:2022rwr}.
The stronger statement that one-gluon-exchange potentials between any constituents of different color-singlet hadrons each vanish individually has been previously noted for meson-meson systems~\cite{Huang:2020dci,Assi:2023dlu}. We are not aware of previous extensions of this result to arbitrary hadron-hadron pairs in QCD or other non-Abelian gauge theories.

Note that vanishing of hadron-hadron van der Waals potentials is only a consequence of generator tracelessness at LO.
In particular, van der Waals potentials are not forbidden by gauge invariance in non-Abelian theories and can arise at higher order; see \cref{sec:NNLO} below.

\subsection{Multipole expansion and Gauss's law}\label{sec:Gauss}

Gauge invariance, and Gauss's law in particular, constrain the multipole expansion of van der Waals potentials at all orders in the gauge coupling expansion.
This section shows how Gauss's law can be used to show that van der Waals potentials vanish at $O(1/R)$ and $O(1/R^2)$ in multipole expansions valid for large hadron-hadron separations $R$.

By “van der Waals potential’’ we mean the inter-hadron piece of the pNRQCD potential, defined by the bookkeeping decomposition
\begin{equation}
V \;=\; V_H \;+\; V_{H'} \;+\; V_{HH'}\,,
\end{equation}
where $V_H$ and $V_{H'}$ act within each color-singlet hadron separately, and $V_{HH'}$ encodes interactions between the two hadrons at separation $R$. We do not change the degrees of freedom or perform an additional matching; $V_{HH'}$ is a subset of terms in the pNRQCD potential, not the potential for a separate van der Waals EFT, in contrast to Refs.~\cite{Brambilla:2015rqa, Brambilla:2017ffe}.

Gauss's law in Yang–Mills theory can be written as
\begin{equation}
D_i^{ab}E^{ib}(x)=g\,\rho^{a}(x),
\end{equation}
where $D_i^{ab}$ is the covariant derivative in the adjoint representation and $E^{ib}$ is the chromoelectric field, constrains the chromoelectric field and therefore potentials that color-singlet states can source.
Any color-singlet hadron state satisfies
\begin{equation}
Q^{a}\,|\Psi\rangle\;=\;\left[\int d^{3}x\,\rho^{a}(x)\right]|\Psi\rangle\;=\;0,
\end{equation}
where $a=1,\dots,d_{\rm Adj}$.
Integrating Gauss's law over a sphere of radius $R$ that encloses the entire hadron and applying Gauss's law gives the vanishing chromoelectric flux condition:
\begin{equation}
\oint_{S_R}dS_{i}\,E^{ia}=g Q^{a}=0,
\end{equation}
for every $R$~\cite{Zwanziger:1998ez, Greensite:2011zz}.

Suppose that a color-singlet hadron with characteristic size $r$ acts as a source for a potential with multipole expansion

\begin{equation}\label{eq:V_multipole}
  \begin{split}
    V^{(\mathbf{1})}(R,r) &= V^{(\mathbf{1},-1)}(\ln(\mu R)) \frac{1}{R} \\
    &\hspace{20pt} + V^{(\mathbf{1},-2)}(\ln(\mu R)) {\ \frac{\bm{r} \cdot \hat{\bm{R}}}{R^2} }\\
    &\hspace{20pt} + O(1/R^3),
  \end{split}
\end{equation}
where the multipole coefficients $V^{(\mathbf{1},-1)},\,V^{(\mathbf{1},-2)},\,\ldots$ can include nonanalytic dependence on $R$ through logarithms but by the definition of the multipole expansion cannot depend on powers of $R$.\footnote{At arbitrary finite loop order, this version of the multipole expansion can be formally viewed as a transseries involving a joint expansion of powers of $1/R$ and of (finitely many) powers of $\ln(\mu R)$ in which all terms involving the same power of $1/R$ are grouped into a coefficient function depending on $\ln(\mu R)$.}
The multipole coefficients may also depend implicitly on $r^2$, $\ln(\mu r)$, and the generators $T^a$.
The corresponding electric field takes the form
\begin{equation}\label{eq:E_multipole}
  \begin{split}
    \bm{E}^{(\mathbf{1})a}(R,r) &= 2\text{Tr}\left[ T^a \left( V^{(\mathbf{1},-1)}(\ln(\mu R)) \vphantom{\frac{\partial V^{(\mathbf{1},-1)}}{\partial \ln(\mu R)}}  \right. \right. \\
    &\hspace{40pt} - \left.\left.  \frac{\partial V^{(\mathbf{1},-1)}}{\partial \ln(\mu R)} \right) \right] \frac{1}{R^2} \hat{\bm{R}} \\
  &\hspace{20pt} + O(1/R^3).
  \end{split}
\end{equation}
Applying Gauss's law then gives
\begin{equation}
  \begin{split}
    0 &= \oint_{S_R}dS_{i}\,E^{(\mathbf{1})ia} \\
    &= 4\pi R^2 \hat{\bm{R}}\cdot \bm{E}^{(\mathbf{1})a}(R,r)  \\
    &= 8\pi \left( \text{Tr}\left[T^a V^{(\mathbf{1},-1)}\right] - \frac{\partial \text{Tr}\left[T^a V^{(\mathbf{1},-1)}\right] }{\partial \ln(\mu R)} \right)  + O(1/R).
  \end{split}
\end{equation}
This can be viewed as a differential equation for $\text{Tr} \left[ T^a V^{(\mathbf{1},-1)} \right]$, which has the solution
\begin{equation}
    \text{Tr} \left[ T^a V^{(\mathbf{1},-1)} \right] = C \exp\left[ \ln(\mu R) \right] = C \mu R,
\end{equation}
where $C$ is a constant.
For any $C \neq 0$, this solution leads to a contradiction with the supposition of the multipole expansion that $V^{(\mathbf{1},-1)}(\ln(\mu R))$ depends logarithmically on $R$ and is not proportional to a power of $R$.\footnote{If this solution were inserted back into the multipole expansion it would lead to an $O(R^0)$ constant term in the potential that would make a vanishing contribution to the chromoelectric field.}
The only valid multipole coefficient solution therefore has $C = 0$ and thus $ \text{Tr} \left[ T^a V^{(\mathbf{1},-1)} \right] = 0$.
This implies that $\bm{E}^{(\mathbf{1})a}(R,r) = O(1/R^3)$ and further that $V^{(\mathbf{1},-1)}$ must be proportional to the $\mathfrak{su}(3)$ identity matrix.

This argument implies 
{\
an analogous cancellation for the potential $V_{HH'}^{(\mathbf{1} \otimes \mathbf{1})}(R,r_1,r_2)$ between two color-singlet hadrons in pNRQCD.
The leading term in the multipole expansion of $V_{HH'}^{(\mathbf{1} \otimes \mathbf{1})}$ is the sum of the potentials $V_H^{(\mathbf{1})}$ and $V_{H'}^{(\mathbf{1})}$ sourced by each hadron.
The $1/R$ terms in each vanish by the Gauss's law arguments above, meaning that the $1/R$ term in the hadron-hadron potential multipole expansion
\begin{equation}\label{eq:V_multipole}
  \begin{split}
    &V_{HH'}^{(\mathbf{1} \otimes \mathbf{1})}(R,r_1,r_2) = V^{(\mathbf{1} \otimes \mathbf{1},-1)}(\ln(\mu R)) \frac{1}{R} \\
    &\hspace{20pt} + V_1^{(\mathbf{1} \otimes \mathbf{1},-2)}(\ln(\mu R)) \left[ \frac{\bm{r}_1 \cdot \hat{\bm{R}}}{R^2} \right] \\
    &\hspace{20pt} + V_2^{(\mathbf{1} \otimes \mathbf{1},-2)}(\ln(\mu R)) \left[  \frac{\bm{r}_2 \cdot \hat{\bm{R}}}{R^2} \right] \\
    &\hspace{20pt} + V^{(\mathbf{1} \otimes \mathbf{1},-3)}_1(\ln(\mu R)) \left[ \frac{(\bm{r}_1 \cdot \hat{\bm{R}}) (\bm{r}_2 \cdot \hat{\bm{R}})}{R^3} \right] \\
    &\hspace{20pt} + V^{(\mathbf{1} \otimes \mathbf{1},-3)}_2(\ln(\mu R)) \left[  \frac{\bm{r}_1 \cdot \bm{r}_2}{R^3} \right]\\
    &\hspace{20pt} + O(1/R^4),
  \end{split}
\end{equation}
vanishes, that is $V^{(\mathbf{1} \otimes \mathbf{1},-1)} = 0$.
}
In weakly-coupled pNRQCD with only gluon and nonrelativistic heavy quark degrees of freedom,\footnote{If light quark degrees of freedom are present, then color-singlet two-meson exchange potentials may be present that fall off exponentially but may take an approximate $1/R$ form for scales smaller than the inverse light meson mass~\cite{Brambilla:2015rqa}.
It is also possible that nonperturbative effects that vanish at all orders in $\alpha$, for instance associated with color-singlet glueball exchange, could lead to contributions to the pNRQCD potential that take an approximate $1/R$ form for scales smaller than $1/\Lambda_{\rm QCD}$.
Further calculations outside the scope of weakly-coupled pNRQCD are needed to investigate whether small but non-zero binding energies (for instance suppressed by powers of $e^{-1/\alpha}$) could result from such effects.} it follows that the potential between color-singlet hadrons vanishes faster than $1/R$.

One can place even stronger bounds than this.
From \cref{eq:V_multipole}, one can {\ further} see that the potential sourced by the first hadron is proportional to its dipole moment $\mathbf{r_1}$, so the leading contribution to the multipole expansion of $V_{HH'}$ must be proportional to $\mathbf{r}_1$. 
{\ This shows that $V_2^{(\mathbf{1} \otimes \mathbf{1},-2)} = 0$.}
Repeating this argument for the second hadron tells us that $V_{HH'}$ must similarly be proportional to $\mathbf{r}_2$,
{\ and so $V_1^{(\mathbf{1} \otimes \mathbf{1},-2)} = 0$.}
Since the potential has total mass dimension $+1$ and the powers of $\mathbf{r}_1, \mathbf{r}_2$ must be cancelled by inverse powers of the inter-hadron separation $R$, the potential must fall off at least as fast as $R^{-3}$ (up to logarithms).

This is consistent with previous studies of quarkonium-quarkonium systems that have studied the same multipole expansion and found that van der Waals potentials are either of dipole-dipole form or involve two-pion exchange diagrams~\cite{Peskin:1979va, Bhanot:1979vb, Brambilla:2015rqa}.
{\
Note that this long-range form is only applicable for scales $1/m_Q \ll R \ll 1/\Lambda_{\rm QCD}$ and does not necessarily coincide with the asymptotic form of the potential that would appear in a van der Waals EFT~\cite{Brambilla:2015rqa,Brambilla:2017ffe} that was matched directly to nonpertubative QCD.
}

This argument applies to the pNRQCD static potential at any finite order in $\alpha$.
It further applies to $O(1/m_Q^{n})$ corrections to the potential for any power $n$.
The only critical assumptions in the argument are that the nonrelativistic expansion underlying pNRQCD is applicable for all quark degrees of freedom; potentials that are nonanalytic in $1/m_Q$, such as $e^{-m_Q R} / R$, taking a monopole form for $R \ll 1/m_Q$, are not excluded by this argument.
In particular, such potentials nonanalytic in $1/m_Q$ include the $e^{-m_\pi R} / R$ Yukawa potentials between color-singlet baryons familiar from nuclear physics.

\subsection{NLO van der Waals potentials cannot bind}\label{sec:nobind}

In general, potentials falling at least as fast as $1/R^2$ do not form bound states if their coefficient is small.
More explicitly, suppose that the intra-hadron potential is $\alpha/r$ so that the characteristic size of individual hadrons scales as $r \sim 1/(\alpha m_Q)$, where here and below $\sim$ denotes proportionality in the small-$\alpha$ limit. Further suppose that the terms depending on the inter-hadron distance $R$ take the form
\begin{equation}\label{eq:piecewise}
    V(R) \sim -\left\{ \begin{matrix} \frac{\alpha^{1+n}}{R} \left(\frac{r}{R}\right)^{1+k} & (R > r) \\ \frac{\alpha^{1+n}}{R} & (R < r) \end{matrix} \right.
\end{equation}
for $k \geq 0, n > 0$.  
Then there are no hadron-hadron bound states whose color state is a product of color-singlets in the $\alpha \rightarrow 0$ limit.

{\
Before proceeding to a proof, note that the pNRQCD potential between two color-singlet hadrons has the same asymptotic behavior as Eq.~\eqref{eq:piecewise}.
The large-$R$ behavior of $V^{(\mathbf{1}\otimes\mathbf{1})}$ is argued to be $O(R^{-3})$ based on the Gauss's law arguments above, which satisfies the constraints $k\geq 0$ and $n>0$ on the large-$R$ form of Eq.~\eqref{eq:piecewise}.
The small-$R$ behavior cannot be stronger than the Coulombic behavior of the pNRQCD potential between hadron constituents.
The constraint that $n>0$ in the $\alpha^{1+n}/R$ form at small $R$ comes from the result in Sec.~\ref{sec:LO} that $V^{(\mathbf{1}\otimes\mathbf{1})}$ vanishes identically at LO and therefore any non-vanishing terms are suppressed by $\alpha^{1+n}$ with $n>0$.
}

To demonstrate this, we need to consider two cases: a bound state with radius $R \lesssim r$ and one with radius $R \gtrsim r$.
In both cases, the uncertainty principle dictates that the momentum variable $P$ conjugate to the intra-radius separation $R$ scales as $P \sim 1 / R$.
In the first case, where $R \lesssim r$ and the bound state is relatively compact, the potential energy associated with $R$ is $-\alpha^{1+n}/R$ and the associated kinetic energy is $P^2/m_Q \sim 1/m_Q R^2$.
The sum of these contributions is
\begin{equation}
  \begin{split}
    \frac{1}{R} \left(\frac{1}{m_Q R} - \alpha^{1+n}\right) &\gtrsim \frac{1}{R} \left(\frac{1}{m_Q r} - \alpha^{1+n}\right) \\
    &\sim \frac{\alpha}{R} (1-\alpha^n) > 0
  \end{split}
\end{equation}
using the fact that $R \lesssim r \sim 1/\alpha$ and $\alpha^n \ll 1$ for $n>0$.
Positivity of the total energy contribution contradicts the existence of a bound state.

In the second case, where $R \gtrsim r$ and the bound state is relatively diffuse, the same sum of kinetic and potential contributions becomes
\begin{equation}
  \begin{split}
    \frac{1}{R} \left(\frac{1}{m_Q R} - \alpha^{1+n} (r/R)^{1+k} \right) &\sim \frac{1}{m_Q R^2} \left(1 - \alpha^{n} (r/R)^{k} \right) \\
    &\gtrsim \frac{1}{m_Q R^2} (1-\alpha^{n}) > 0,
  \end{split}
\end{equation}
where the first line uses $r \sim 1/(\alpha m_Q)$ and the inequalities in the second line follow from $r/R \lesssim 1$ and $\alpha \ll 1$.
Again, the total energy is positive, so there is no bound state.

Note that this theorem does not apply to a potential of the form $\log(\mu R)/R^2$, which falls off more slowly than $1/R^2$.  However, it does apply to $\ln^p (\mu R)/R^3$ for any power $p$.

The results of \cref{sec:LO,sec:Gauss} imply that pNRQCD van der Waals potentials between color-singlet hadrons are $O(\alpha^2 \ln^p(\mu R)/R^3)$ and therefore satisfy the theorem with $n=1$ and $k=0$.
This in turn implies that there are no hadron-hadron bound states whose color state is a product of singlets in $SU(N)$, $SO(N)$, or $Sp(2N)$ gauge theories in the limit where all quark masses satisfy $m_Q \gg \Lambda_{\rm QCD}$. We reiterate that this ignores long-range forces from light-pion exchange, as well as nonperturbative corrections to weakly-coupled pNRQCD, and thus does not entirely describe finite-mass heavy flavors such as $b$ quarks in real QCD.

\section{Exotic color configurations in pNRQCD at NLO}\label{sec:NLO}

The results of the previous section imply that color-singlet-product hadron-hadron bound states do not exist when $m_Q \gg \Lambda_{\rm QCD}$ but leave open the possibility of hadron-hadron bound states whose color structure is more exotic than a product of singlets.
We do not know of a general way to prove whether or not such bound states exist analytically.
Instead, the energies of different color configurations must be computed numerically for particular gauge groups.
Although it is impossible to cover the infinitely large Hilbert space of space-color configurations, it is possible to use pNRQCD to scan over a wide range of color configurations with physically motivated spatial wavefunctions.

This section explores this larger space of color configurations for $SU(3)$ gauge theory with $m_Q \gg \Lambda_{\rm QCD}$ using GFMC calculations of four-, five-, and six-quark systems.
First, the pNRQCD formalism, valid through NNLO and GFMC methods, is briefly reviewed.
Complete bases of tensors and physically motivated trial wavefunctions are then constructed for color-singlet tetraquark, pentaquark, and hexaquark states.

\subsection{pNRQCD}
\label{sec:pnrqcd}

The pNRQCD Hamiltonian is given by
\begin{equation}
  \Ham = \Top + \Vop^{\psi\chi} + \Vop^{\psi\psi} + \Vop^{\chi\chi} + \ldots,
\end{equation}
where $\Top$ is the nonrelativistic kinetic energy operator defined in terms of heavy quark fields $\psi(\bs{r})$ and antiquark fields $\chi(\bs{r})$ by 
\begin{equation}
\Top = -\int d^3 \bs{r}\ \left[\psi_i^\dagger(\bs{r}) \frac{\nabla^2}{2m_Q} \psi_i(\bs{r}) + \chi_i^\dagger(\bs{r}) \frac{\nabla^2}{2m_Q} \chi_i(\bs{r})\right],
\end{equation}
where $i$ is a color index, and two-component spinor indices are suppressed.
The potential terms in $H$ are computed in a joint expansion in powers of $1/m_Q$, and the strong coupling constant $\alpha_s$ is evaluated at scales proportional to $m_Q$. In what follows, we count each insertion of $1/m_Q$ as the same order as a power of $\alpha_s(\mu_p)$ where $\mu_p \sim \alpha_s m_Q$, in line with the joint NRQCD/pNRQCD power‑counting prescriptions of Refs.~\cite {Bodwin:1994jh,Brambilla:1999xf}.

The quark-antiquark potential operator $\Vop^{\psi\chi}$ is 
\begin{equation}
    \begin{split}
        V^{\psi\chi} &= \int d^3\bs{r}_1d^3\bs{r}_2\,  \psi^{\dagger}_i(\bs{r}_1)\chi_j(\bs{r}_2) \chi^{\dagger}_k(\bs{r}_2)\psi_l(\bs{r}_1) \\
       &\hspace{10pt} \times \left[ \frac{1}{3}\delta_{ij}\delta_{kl} V^{\psi\chi}_{\mathbf{1}}(\bs{r}_{12})  + 2 T^a_{ji} T^a_{kl} V^{\psi\chi}_{\text{Adj}}(\bs{r}_{12}) \right], \label{eq:Lpsichi}
    \end{split}
\end{equation}
where the $T^a$ are $\mathfrak{su}(3)$ generators normalized as $\text{Tr}[T^a T^b] = \frac{1}{2} \delta^{ab}$.
The $V^{\psi\chi}_{\mathbf{1}}(\bs{r}_{12})$ and  $V^{\psi\chi}_{\text{Adj}}(\bs{r}_{12})$ are color-singlet and color-adjoint potentials respectively.
It is convenient to define the color-singlet potential as~\cite{brodsky:1999fr}
\begin{equation}
\begin{aligned}
  V^{\psi\chi}_{\mathbf{1}}(\bs{r}) \equiv -\frac{4\alpha_V}{3|\bm{r}|},
\end{aligned}
\end{equation}
where the tree-level Casimir color factor $C_F = 4/3$ entering the $Q\overline{Q}$ one-gluon-exchange diagram is factored out and $\alpha_V = \alpha_s + O(\alpha_s^2)$ is defined to absorb the loop-level corrections to the potential that were computed to NNLO in Ref.~\cite{Schroder:1998vy} and N$^3$LO in~\cite{Anzai:2013tja,Lee:2016cgz}. 
When computed in fixed-order perturbation theory, $\alpha_V$ and therefore $V^{\psi\chi}_{\mathbf{1}}(\bs{r})$ are renormalization scale dependent and include factors of $\ln(\mu |\bm{r}|)$, as defined in Ref.~\cite{Assi:2023dlu}. We are currently incorporating the renormalon subtracted and resummed static potential in the Minimal Renormalon Subtracted (MRS) scheme~\cite{Kronfeld:2024qao,Brambilla:2017hcq}; results will be presented in forthcoming work~\cite{Assi:inprep}.

The adjoint potential can then be defined as
\begin{equation}
\begin{aligned}
    V^{\psi\chi}_{\mathbf{8}}(\bs{r}) = \frac{\alpha_V}{6|\bm{r}|} + \frac{\delta a_2^{\mathbf{8}}}{|\bm{r}|},
\end{aligned}
\end{equation}
where the tree-level color factor is again isolated and $\delta a_2^{\mathbf{8}}$ is defined to absorb representation-dependent loop-level corrections to the potential. These corrections first appear at NNLO and were computed in Ref.~\cite{Collet:2011kq,Kniehl:2004rk} as 
\begin{equation}\label{eq:a2r8}
  \delta a_2^{\mathbf{8}} = \alpha_s \left( \frac{\alpha_s}{4\pi} \right)^2 \left[ -9 \pi^2 (12 - \pi^2) \right] + O(\alpha_s^4).
\end{equation}
The quark-quark potential is given by 
      \begin{align}\label{eq:Lpsipsi}
        &V^{\psi\psi} = \int d^3\bs{r}_1d^3\bs{r}_2\,  \psi^{\dagger}_i(\bs{r}_1)\psi^{\dagger}_j(\bs{r}_2) \psi_k(\bs{r}_2)\psi_l(\bs{r}_1) \\\nonumber
      &\times \left[ \frac{1}{4} \epsilon_{ijo}\epsilon_{klo}  V^{\psi\psi}_{\overline{\mathbf{3}}}(\bs{r}_{12})  +  \frac{1}{4}  \left(  \delta_{il}\delta_{jk}+\delta_{jl}\delta_{ik}\right)   V^{\psi\psi}_{\mathbf{6}}(\bs{r}_{12}) \right],
    \end{align}   
where $V^{\psi\psi}_{\overline{\mathbf{3}}}(\bs{r}_{12})$ and $V^{\psi\psi}_{\mathbf{6}}(\bs{r}_{12})$ involve color-antisymmetric and color-symmetric products of quark fields, respectively.
The color-antisymmetric $QQ$ potential is given by
\begin{equation}
\begin{aligned}
  V^{\psi\psi}_{\overline{\mathbf{3}}}(\bs{r}) = -\frac{2\alpha_V}{3|\bm{r}|} + \frac{\delta a_2^{\overline{\mathbf{3}}}}{|\bm{r}|},
\end{aligned}
\end{equation}
where representation-dependent differences besides the tree-level color factor again arise at NNLO and were computed in  Ref.~\cite{Assi:2023cfo} to be
\begin{equation}
  \delta a_2^{\overline{\mathbf{3}}} = \alpha_s \left( \frac{\alpha_s}{4\pi} \right)^2 \left[ -\frac{3}{2} \pi^2 (12 - \pi^2) \right] + O(\alpha_s^4).
\end{equation}
The color-symmetric $QQ$ potential is given similarly to
\begin{equation}
\begin{aligned}
  V^{\psi\psi}_{\mathbf{6}}(\bs{r}) = \frac{\alpha_V}{6|\bm{r}|} + \frac{\delta a_2^{\mathbf{6}}}{|\bm{r}|},
\end{aligned}
\end{equation}
where~\cite{Assi:2023cfo}
\begin{equation}
  \delta a_2^{\mathbf{6}} = \alpha_s \left( \frac{\alpha_s}{4\pi} \right)^2 \left[ -\frac{15}{2} \pi^2 (12 - \pi^2) \right] + O(\alpha_s^4).
\end{equation}
The antiquark-antiquark potential is identical by charge conjugation, and $V^{\chi\chi}$ is obtained from Eq.~\eqref{eq:Lpsipsi} via the replacement $\psi \rightarrow \chi$.
Three- and four-quark potentials enter $\Ham$ at NNLO~\cite{Brambilla:2009cd,Assi:2023cfo}.
Moreover, spin-independent and spin–dependent interactions first arise at ${\cal O}(\alpha_s^2/m_Q)$ and ${\cal O}(\alpha_s/m_Q^{2})$, respectively, and contribute to the spectrum already at NNLO~\cite{Brambilla:2000gk,Brambilla:2004jw,assi:2020piz}. Effects from ultra-soft modes lead to the appearance of additional non-potential terms in $\Ham$, but these do not enter until N${}^3$LO~\cite{Pineda:1997bj,Pineda:1998kn,Kniehl:1999ud,Brambilla:1999xj,Manohar:2000kr}.

The eigenvalues of $\Ham$, denoted $\Delta E$ below, are the masses of pNRQCD energy eigenstates minus the rest masses $m_Q$ of their constituent heavy quarks/antiquarks.
The definition of $m_Q$ and choice of renormalization scheme and scale $\mu$ will modify $\Delta E(m_Q,\mu)$ such that the masses of pNRQCD energy eigenstates are scheme- and scale-independent up to perturbative truncation effects.
We use the quark mass scheme and results for $m_b$ and $m_c$ obtained in Ref.~\cite{Assi:2023cfo} in which the ``pole masses'' $m_Q$ appearing in $\Ham$ are obtained by solving
\begin{equation}
    M_{Q\overline{Q}} = 2 m_Q + \Delta E_{Q\overline{Q}},
\end{equation}
using experimental results for $M_{Q\overline{Q}}$,
where  $\Delta E_{Q\overline{Q}}(m_Q, \mu)$ is computed using $\mu = \mu_p(m_Q)$ obtained numerically as the solution to
\begin{equation}
     \mu_p(m_Q)\equiv 4 \alpha_s(\mu_p(m_Q)) m_Q. \label{eq:mup}
\end{equation}

Heavy quarkonium states only involve the color-singlet $\psi\chi$ potential, which is attractive and leads to a hydrogen-like spectrum of $Q\overline{Q}$ bound states.
Triply-heavy baryon states only involve the color-antisymmetric $\psi\psi$ potential, which is attractive and leads to the appearance of $QQQ$ bound states.
Color-singlet with four or more heavy quarks/antiquarks, which include multihadron states as well as possibly bound tetraquarks, pentaquarks, hexaquarks, and more, involve the complete set of attractive and repulsive potentials described above.

\subsection{Quantum Monte Carlo}
\label{sec:qmc}

For an arbitrary trial state $\ket{\Psi_T(\bs{\omega})}$ with parameters $ \bs{\omega} = (\omega_1,\ldots) $, the variational principle dictates that $\Delta E \leq \mbraket{\Psi_T(\bs{\omega})}{\Ham}{\Psi_T(\bs{\omega})}$.
Numerical minimization of $\mbraket{\Psi_T(\bs{\omega})}{\Ham}{\Psi_T(\bs{\omega})}$ can therefore be used to determine the best ground-state approximation within a parameterized family of wavefunctions~\cite{toulouse2007optimization}.
We evaluate these matrix elements using 
wavefunctions $\Psi_T(\bs{R};\bs{\omega}) \equiv \braket{\bs{R}}{\Psi_T(\bs{\omega})}$ that depend on spatial coordinates $\bs{R} \equiv (\bs{r}_1,\ldots,\bs{r}_{N_Q})$,
\begin{equation}
  \mbraket{\Psi_T}{\Ham}{\Psi_T}  = \frac{\int d^3\bs{R}\, \Psi_T(\bs{R})^*H(\bs{R}) \Psi_T(\bs{R})}{\int d^3\bs{R}\, |\Psi_T(\bs{R})|^2 }, \label{eq:me}
\end{equation}
where $ \mbraket{\bs{R}}{\Ham}{\bs{R}'} = H(\bs{R}) \delta(\bs{R}-\bs{R}') $, states are assumed to be normalized as $\braket{\Psi_T}{\Psi_T} = 1$, and dependence on $\bs{\omega}$ is suppressed for brevity.
We  use Monte Carlo methods to stochastically approximate Eq.~\eqref{eq:me} by sampling $\bs{R}$ from a probability distribution proportional to $|\Psi_T(\bs{R})|^2$ and then obtaining $\mbraket{\Psi_T}{H}{\Psi_T}$ as the sample mean of $\Psi_T(\bs{R})^*H(\bs{R}) \Psi_T(\bs{R})$ for this ensemble.

The accuracy of ground-state determinations using this variational Monte Carlo (VMC) approach is limited by the expressivity of a given family of trial wavefunctions, and we therefore adopt the standard QMC strategy of using optimal trial wavefunctions obtained using VMC as the foundation for subsequent GFMC calculation~\cite{Carlson:2014vla,Gandolfi:2020pbj}. GFMC employs imaginary-time $\tau$ evolution (analogous to lattice QCD calculations) to dampen the excited-state components of $\ket{\Psi_T}$ and formally allows the ground-state for a set of quantum numbers to be obtained from any trial wavefunction with the same quantum numbers as $\lim_{\tau \to \infty} e^{-\Ham \tau}\ket{\Psi_T}$.
The imaginary-time evolution operator $e^{-\Ham\tau}$ cannot be straightforwardly constructed for arbitrary $\tau$, but it can be approximated by splitting $\tau$ into $N_\tau$ intervals of size $\delta\tau=\tau/N_\tau$ for $N_\tau \gg 1$ and using the Lie-Trotter product formula:
\begin{align}
    \mbraket{\bs{R}}{e^{-\Ham\tau}}{\Psi_T} \approx {}&\int\prod_{i=0}^{N_{\tau}-1}d\bs{R}_i\bra{\bs{R}_{N_{\tau}}}e^{-H\delta\tau}\ket{\bs{R}_{N-1}}\times \cdots \nonumber\\{}&
    \times \bra{\bs{R}_1}e^{-H\delta\tau}\ket{\bs{R}_{0}}\braket{\bs{R}_0}{\Psi_T},
\end{align}
with equality obtained in the $N_\tau \rightarrow \infty$ limit.
The Green's functions $G_{\delta\tau}(\bs{R},\bs{R}' ) \equiv \mbraket{\bs{R}}{e^{-\Ham\delta\tau}}{\bs{R}'}$ are approximated with the Trotter-Suzuki expansion 
\begin{equation}
\begin{split}
    &G_{\delta\tau}(\bs{R},\bs{R}' ) \equiv \mbraket{\bs{R}}{e^{-\Ham\delta\tau}}{\bs{R}'} \\
    &\hspace{20pt} \approx e^{-V(\bs{R})\delta \tau/2} \mbraket{\bs{R}}{e^{-T \delta\tau}}{\bs{R}'} e^{-V(\bs{R}')\delta \tau/2},
    \end{split}\label{eq:VK}
\end{equation}
where the kinetic piece $\mbraket{\bs{R}}{e^{-T \delta\tau}}{\bs{R}'}$ is proportional to a Gaussian $e^{-(\bs{R}-\bs{R}')^2/\lambda^2}$ with $\lambda^2 = 2 \delta \tau / m_Q$~\cite{Carlson:2014vla,Gandolfi:2020pbj}.
Therefore, GFMC evolution for each Trotter step can be achieved by sampling $\bs{R} - \bs{R}'$ from a Gaussian distribution and computing the action of the potential on coordinate-space states.
We further employ strategies to improve the precision of GFMC by randomly choosing between updates with $\pm(\bs{R} - \bs{R}')$ as detailed in Ref.~\cite{Gandolfi:2020pbj}.
The kinetic piece is diagonal in color, while for a state built from $N_Q$ heavy quark/antiquark fields, the potential is represented as a $3N_Q \times 3N_Q$ color matrix, and we approximate the matrix exponentials in Eq.~\eqref{eq:VK} using a second-order Taylor expansion.

Calculations of Hamiltonian matrix elements after imaginary-time evolution provide effective energies
\begin{equation}
  \Delta E(\tau) = \frac{ \mbraket{\Psi_T}{H e^{-H \tau}}{\Psi_T} }{ \mbraket{\Psi_T}{ e^{-H \tau}}{\Psi_T} }.
\end{equation}
These effective energies approach the ground-state energies $\Delta E$ in the $\tau \rightarrow \infty$ limit and include additional exponentially-suppressed contributions from excited states at finite $\tau$ as summarized in Ref.~\cite{Assi:2023cfo}.
The contributions are necessarily positive, and therefore $\Delta E(\tau)$ provides a variational bound that must approach the ground-state energy from above.

\subsection{Tetraquarks / meson-meson systems}

The pNRQCD potentials for $\overline{Q}Q\overline{Q}Q$ states, which can describe bound tetraquarks or unbound meson-meson systems, at LO and NLO are discussed in Refs.~\cite{Czarnecki:2017vco,Assi:2023dlu} and reproduced here for completeness.
A convenient basis of interpolating operators for these states is
\begin{equation} \label{eq:4Q_O_def}
  \mathcal{O}^{(\overline{Q}Q\overline{Q}Q,\mathcal{C})} = \chi_i \psi_j \chi_k \psi_l \mathcal{T}^{(\overline{Q}QQQQ,\mathcal{C})}_{ijkl},
\end{equation}
where $\mathcal{C} \in \{ \overline{\mathbf{3}}\otimes\mathbf{3},\  \mathbf{6}\otimes\overline{\mathbf{6}} \}$ labels the two ways of combining diquarks (color representations $\mathbf{3}\otimes\mathbf{3} = \overline{\mathbf{3}}\oplus \mathbf{6}$)  and antidiquarks (color representations $\overline{\mathbf{3}}\otimes\overline{\mathbf{3}} = \mathbf{3}\oplus \overline{\mathbf{6}}$) into color singlets.
The color tensors for these operators are given by
\begin{equation}
  \begin{aligned} \label{eq:4Q_1x1_def}
  \mathcal{T}^{(\overline{Q}Q\overline{Q},\overline{\mathbf{3}}\otimes\mathbf{3})} &= \frac{1}{2\sqrt{3}} \left(\delta_{ij}\delta_{kl} - \delta_{il}\delta_{jk} \right)., \\
  \mathcal{T}^{(\overline{Q}Q\overline{Q},\mathbf{6}\otimes\overline{\mathbf{6}})} &= \frac{1}{2\sqrt{6}} \left(\delta_{ij}\delta_{kl} + \delta_{il}\delta_{jk} \right).
\end{aligned}
\end{equation}
The potentials for coordinate-basis states $\ket{\overline{Q}Q\overline{Q}Q} \equiv \ket{\chi(\bm{x}_1) \psi(\bm{x}_2) \chi(\bm{x}_3) \psi(\bm{x}_3)}$ are defined by the matrix elements
\begin{equation}\label{eq:potential_coords}
  V_{\mathcal{C}}^{(\overline{Q}Q\overline{Q}Q)} \equiv \mbraket{ \overline{Q}Q\overline{Q}Q  }{ V^{\psi\chi} + V^{\psi\psi} }{  \overline{Q}Q\overline{Q}Q}.
\end{equation}
For these color basis states, they are given up to NLO by~\cite{Huang:2020dci,Assi:2023dlu}
\begin{equation}\label{eq:4Q_3barx3}
\begin{split}
    \frac{V_{\overline{\mathbf{3}}\otimes\mathbf{3}}^{(\overline{Q}Q\overline{Q}Q,\text{NLO})}}{\alpha_V} &= -\frac{2}{3} \left( \frac{1}{|\bm{r}_{13}|} + \frac{1}{|\bm{r}_{24}|} \right) \\
    &\hspace{20pt} - \frac{1}{3} \left( \frac{1}{|\bm{r}_{12}|} + \frac{1}{|\bm{r}_{14}|} + \frac{1}{|\bm{r}_{23}|}  + \frac{1}{|\bm{r}_{34}|}  \right) ,
    \end{split}
\end{equation}
where $\bm{r}_{ij} \equiv \bm{r}_i - \bm{r}_j$,  and
\begin{equation}\label{eq:4Q_6x6bar}
\begin{split}
  \frac{V_{\mathbf{6}\otimes\overline{\mathbf{6}}}^{(\overline{Q}Q\overline{Q}Q,\text{NLO})}}{\alpha_V} &= \frac{1}{3} \left( \frac{1}{|\bm{r}_{13}|} + \frac{1}{|\bm{r}_{24}|} \right) \\
    &\hspace{20pt} - \frac{5}{6} \left( \frac{1}{|\bm{r}_{12}|} + \frac{1}{|\bm{r}_{14}|} + \frac{1}{|\bm{r}_{23}|}  + \frac{1}{|\bm{r}_{34}|}  \right) .
    \end{split}
\end{equation}
LO results are immediately obtained by taking $\alpha_V = \alpha_s + O(\alpha_s^2)$; NNLO results are non-trivially different as discussed below.

{\
The color tensors $\mathcal{C} \in \{\overline{\mathbf{3}}\otimes\overline{\mathbf{3}}, \mathbf{6}\otimes \mathbf{6}\}$ provide a complete basis for $\overline{Q}Q\overline{Q}Q$ color wavefunctions.
In particular, alternative structures involving products of color-singlet and color-adjoint structures can be written as linear combinations of these two basis tensors using the identity $T^a_{ij}T^a_{kl}=\tfrac12(\delta_{il}\delta_{jk}-\tfrac13\delta_{ij}\delta_{kl})$.
Even so,
}
it is convenient to extend the basis above to an overcomplete set including an operator with $\mathcal{C} = \mathbf{1}\otimes\mathbf{1}$ describing a product of color-singlet mesons
\begin{equation}
  \mathcal{T}^{(\overline{Q}Q\overline{Q},\mathbf{1}\otimes\mathbf{1})} = \frac{1}{3} \delta_{ij}\delta_{kl}.
\end{equation}
As discussed in Refs.~\cite{Huang:2020dci,Assi:2023dlu} and \cref{sec:LO}, the potential for this color state is simply the sum of the two intra-meson potentials,
\begin{equation}\label{eq:4Q_V1x1}
\begin{split}
    \frac{V_{\mathbf{1}\otimes\mathbf{1}}^{(\overline{Q}Q\overline{Q}Q,\text{NLO})}}{\alpha_V} &= -\frac{4}{3} \left( \frac{1}{|\bm{r}_{12}|} + \frac{1}{|\bm{r}_{34}|} \right),
    \end{split}
\end{equation}
with vanishing van der Waals potentials involving constituents of different mesons at LO and NLO.

The vanishing of van der Waals potentials implies that color-singlet meson-meson product states are exact eigenstates of the pNRQCD Hamiltonian at LO and NLO.
However, this does not immediately imply that they are the lowest-energy eigenstates.
To determine whether tetraquark-bound states are present in pNRQCD, Hamiltonian matrix elements must be evaluated for a complete basis of color states.
This calculation was performed at LO in Ref.~\cite{Czarnecki:2017vco} and extended to NLO in Ref.~\cite{Assi:2023dlu} with the result\footnote{Ref.~\cite{Anwar:2017toa} obtained the opposite result at LO; the trial wavefunction claimed to lead to equal-mass bound states in that work was explicitly studied in Ref.~\cite{Assi:2023dlu} and found not to lead to an equal-mass bound tetraquark state.}
that the minimum-energy linear combination of $\overline{\mathbf{3}}\otimes\mathbf{3}$ and $\mathbf{6}\otimes\overline{\mathbf{6}}$ color states is equivalent to the $\mathbf{1}\otimes\mathbf{1}$ state.
One caveat must be noted: although the Hilbert space of (normalized) tetraquark color states is one-dimensional, the Hilbert space of spatial configurations is infinite-dimensional and cannot be searched exhaustively.

The numerical explorations in Refs.~\cite{Assi:2023dlu} use a variety of spatial wavefunctions, including Hylleraas-type molecular wavefunctions for $\mathbf{1}\otimes\mathbf{1}$, 
\begin{equation}\label{eq:PsiH}
    \Psi_H = e^{-(|\bm{r}_{12}| + |\bm{r}_{34}|)/a} e^{-(|\bm{r}_{13}| + |\bm{r}_{14}| + |\bm{r}_{23}| + |\bm{r}_{24}|)/b},
\end{equation}
where $a = 2/(4/3 \alpha_V)$ is the quarkonium Bohr radius and $b/a$ is a free ``variational'' parameter to be varied.
These are analogous to the variational wavefunction first used to identify the positronium molecule~\cite{Hylleraas:1947zza}, where the minimum-energy configuration has $b/a \approx 5.8$.
For pNRQCD, the minimum-energy configuration is given by the uncorrelated meson product wavefunction corresponding to $b/a \rightarrow \infty$, indicating that there is no meson molecule bound state analogous to the positronium molecule.

Similar wavefunctions are used for other color configurations with $4/3$ replaced by the color factor $\mathcal{C}_{ij}$ multiplying $-1/|\bm{r}_{ij}|$ in the potential for that color configuration,
\begin{equation}\label{eq:PsiH_general}
    \Psi_H = \prod_{i<j} e^{-|\bm{r}_{ij}| / a_{ij}},
\end{equation}
where $a_{ij} \equiv 2 / (\mathcal{C}_{ij} \alpha_V)$ and the product runs over all pairs of indices where $\mathcal{C}_{ij} > 0$, corresponding to attractive interactions.

These numerical simulations cannot formally prove that a more exotic configuration does not lead to a bound state.
However, the pNRQCD potential is a smooth function of the spatial coordinates. These numerical results scanning over the whole Hilbert space of color states with physically motivated spatial wavefunctions and evolving for large imaginary times $\tau \gg 1/(\alpha_s^2 m_Q)$ provide strong evidence that $\mathbf{1}\otimes\mathbf{1}$  states are the lowest energy configuration of equal-mass $\overline{Q}Q\overline{Q}Q$ systems at LO and NLO.

The situation for unequal-mass $\overline{Q}Q\overline{Q}Q$ states is more complicated.
For $\overline{b}t\overline{b}t$ states with $m_t \gg m_b$ for example, the
$\overline{\mathbf{3}}\otimes\mathbf{3}$ configuration with a compact $tt$ diquark ``core'' surrounded by a much less compact $\overline{b}\overline{b}$ antidiquark ``cloud'' leads to a lower-energy configuration describing a bound unequal-mass tetraquark state at LO~\cite{Czarnecki:2017vco,Assi:2023dlu} and NLO~\cite{Assi:2023dlu}.

The non-existence of equal-mass tetraquarks and binding of unequal mass tetraquarks can both be reproduced by a straightforward prescription: approximate the total energy as the independent sum of Coulomb binding energies $-(2m_i m_j)/(m_i+m_j) \mathcal{C}_{ij}^2 \alpha_V^2 / 4$ for each attractive potential $\alpha_V \mathcal{C}_{12}/|\bm{r}_{ij}|$ between quarks with masses $m_i$ and $m_j$.
Although this naive approximation is not quantitatively valid in pNRQCD, it correctly reproduces the relative ordering of variational energies for $\mathbf{6}\otimes\overline{\mathbf{6}}$, $\overline{\mathbf{3}}\otimes\mathbf{3}$, and $\mathbf{1}\otimes\mathbf{1}$ color states in the equal-mass case and even quantitatively reproduces the binding energies of unequal-mass tetraquarks in the extreme mass-ratio limit~\cite{Assi:2023dlu}.

\subsection{Hexaquarks / baryon-baryon systems}

Six-quark, $6Q$, systems can be analyzed analogously using interpolating operators of the form
\begin{equation}\label{eq:6Q_O_def}
    \mathcal{O}^{(6Q,\mathcal{C})} = \psi_i \psi_j \psi_k \psi_l \psi_m \psi_n \mathcal{T}^{(6Q,\mathcal{C})}_{ijklmn}.
\end{equation}
There are five independent ways to form color singlets from a product of six $\mathbf{3}$ representations, which can be conveniently described as products of three diquarks and classified by whether each diquark is antisymmetric ($\overline{\mathbf{3}}$) or symmetric ($\mathbf{6}$).
The combinations that give rise to color singlets are $\mathcal{C} \in \{AAA, AAS, ASA, SAA, SSS\}$, with color tensors given by~\cite{Detmold:2024iwz}
\begin{equation}
    \begin{aligned}
        \mathcal{T}^{(6Q,AAA)}_{ijklmn} &= \varepsilon_{ijm}\varepsilon_{kln}-\varepsilon_{ijn}\varepsilon_{klm},\\
        \mathcal{T}^{(6Q,AAS)}_{ijklmn} &= \varepsilon_{ijm}\varepsilon_{kln}+\varepsilon_{ijn}\varepsilon_{klm},\\
        \mathcal{T}^{(6Q,SSS)}_{ijklmn} &= \varepsilon_{ikm}\varepsilon_{jln}+\varepsilon_{ikn}\varepsilon_{jlm}\\
        &\hspace{20pt} +\varepsilon_{jkm}\varepsilon_{iln}+\varepsilon_{jkn}\varepsilon_{ilm},
    \end{aligned}
\end{equation}
with $\mathcal{T}^{(6Q,ASA)}_{ijklmn}$ and $\mathcal{T}^{(6Q,SAA)}_{ijklmn}$ defined by $\mathcal{T}^{(6Q,AAS)}_{ijklmn}$ with the permutations $(i,j,k,l,m,n) \rightarrow (i,j,m,n,k,l)$ and $(i,j,k,l,m,n) \rightarrow (m,n,i,j,k,l)$, respectively.

The potentials for coordinate basis states $\ket{6Q} \equiv \ket{\psi(\bm{x}_1) \psi(\bm{x}_2) \psi(\bm{x}_3) \psi(\bm{x}_3) \psi(\bm{x}_4) \psi(\bm{x}_5) \psi(\bm{x}_6)}$ can again be defined exactly as in \cref{eq:potential_coords},
\begin{equation}
  V_{\mathcal{C}}^{(6Q)} \equiv \mbraket{ 6Q  }{  V^{\psi\psi} }{ 6Q}.
\end{equation}
They are given for the $AAA$ color state by
\begin{equation}\label{eq:6Q_AAA}
\begin{split}
    \frac{V_{AAA}^{(6Q,\text{NLO})}}{\alpha_V} &= -\frac{2}{3} \left( \frac{1}{|\bm{r}_{12}|} + \frac{1}{|\bm{r}_{34}|} + \frac{1}{|\bm{r}_{56}|}  \right) \\
    &\hspace{10pt} - \frac{1}{6} \left( \frac{1}{|\bm{r}_{13}|} + \frac{1}{|\bm{r}_{14}|} + \frac{1}{|\bm{r}_{15}|} + \frac{1}{|\bm{r}_{16}|}  \right) \\
    &\hspace{10pt} - \frac{1}{6} \left( \frac{1}{|\bm{r}_{23}|} + \frac{1}{|\bm{r}_{24}|} + \frac{1}{|\bm{r}_{25}|} + \frac{1}{|\bm{r}_{26}|}  \right) \\
    &\hspace{10pt} - \frac{1}{6} \left( \frac{1}{|\bm{r}_{35}|} + \frac{1}{|\bm{r}_{36}|} + \frac{1}{|\bm{r}_{45}|}  + \frac{1}{|\bm{r}_{46}|}    \right).
    \end{split}
\end{equation}
This color state is noteworthy for having entirely attractive interactions.
It can be described as three diquark pairs with strongly attractive interactions, where the diquarks have weaker attractive interactions amongst themselves.
The potential for the $SSS$ color state is
\begin{equation}
\begin{split}
    \frac{V_{SSS}^{(6Q,\text{NLO})}}{\alpha_V} &= \frac{1}{3} \left( \frac{1}{|\bm{r}_{12}|} + \frac{1}{|\bm{r}_{34}|} + \frac{1}{|\bm{r}_{56}|}  \right) \\
    &\hspace{10pt} - \frac{5}{12} \left( \frac{1}{|\bm{r}_{13}|} + \frac{1}{|\bm{r}_{14}|} + \frac{1}{|\bm{r}_{15}|} + \frac{1}{|\bm{r}_{16}|}  \right) \\
    &\hspace{10pt} - \frac{5}{12} \left( \frac{1}{|\bm{r}_{23}|} + \frac{1}{|\bm{r}_{24}|} + \frac{1}{|\bm{r}_{25}|} + \frac{1}{|\bm{r}_{26}|}  \right) \\
    &\hspace{10pt} - \frac{5}{12} \left( \frac{1}{|\bm{r}_{35}|} + \frac{1}{|\bm{r}_{36}|} + \frac{1}{|\bm{r}_{45}|}  + \frac{1}{|\bm{r}_{46}|}    \right),
    \end{split}
\end{equation}
which includes a mix of attractive and repulsive potentials.
For the $AAS$ color state, the potential is
\begin{equation}
\begin{split}
    \frac{V_{AAS}^{(6Q,\text{NLO})}}{\alpha_V} &= -\frac{2}{3} \left( \frac{1}{|\bm{r}_{12}|} + \frac{1}{|\bm{r}_{34}|}  \right)  + \frac{1}{3|\bm{r}_{56}|}  \\
    &\hspace{10pt} + \frac{1}{12} \left( \frac{1}{|\bm{r}_{13}|} + \frac{1}{|\bm{r}_{14}|}   \right) - \frac{5}{12} \left( \frac{1}{|\bm{r}_{15}|} + \frac{1}{|\bm{r}_{16}|}  \right)  \\
    &\hspace{10pt} + \frac{1}{12} \left( \frac{1}{|\bm{r}_{23}|} + \frac{1}{|\bm{r}_{24}|}  \right) - \frac{5}{12} \left(  \frac{1}{|\bm{r}_{25}|} + \frac{1}{|\bm{r}_{26}|}  \right)   \\
    &\hspace{10pt} - \frac{5}{12} \left( \frac{1}{|\bm{r}_{35}|} + \frac{1}{|\bm{r}_{36}|}  \right) - \frac{5}{12} \left(  \frac{1}{|\bm{r}_{45}|}  + \frac{1}{|\bm{r}_{46}|}    \right) .
    \end{split}
\end{equation}
The $ASA$ and $SAA$ potentials are related to those of $AAS$ via the permutations described above.
As in the meson-meson and meson-baryon cases, the sum of all color factors is identical for each of the different color states. 

As in the $\overline{Q}Q\overline{Q}Q$ case, products of two color-singlet {\ (or two color-adjoint)} baryons are not linearly independent from the basis tensors described above, and it is convenient to consider an overcomplete set of operators in which {\ color-singlet products} are included.
The LO and NLO potentials for products of two color-singlet baryons, defined by the color tensor
\begin{equation}\label{eq:6Q_1x1_def}
  \mathcal{T}^{(6Q,\mathbf{1}\otimes\mathbf{1})}_{ijklmn} = \frac{1}{6} \varepsilon_{ijk} \varepsilon_{lmn},
\end{equation}
exhibit the same remarkable cancellation of all van der Waals potentials seen above,
\begin{equation}\label{eq:6Q_1x1}
\begin{split}
    \frac{V_{\mathbf{1}\otimes\mathbf{1}}^{(6Q,\text{NLO})}}{\alpha_V} &= -\frac{2}{3} \left( \frac{1}{|\bm{r}_{12}|} + \frac{1}{|\bm{r}_{13}|} + \frac{1}{|\bm{r}_{23}|}  \right) \\
    &\hspace{20pt} - \frac{2}{3} \left( \frac{1}{|\bm{r}_{45}|} + \frac{1}{|\bm{r}_{46}|} + \frac{1}{|\bm{r}_{56}|}  \right) .
    \end{split}
\end{equation}
Therefore, a product state of two color-singlet baryons is an exact eigenstate of the pNRQCD Hamiltonian at LO and NLO.
The question of whether bound hexaquark states exist is whether or not this eigenstate is the ground state with these quantum numbers.

It is noteworthy that the sum of the coefficients of each term in \cref{eq:6Q_AAA}-\cref{eq:6Q_1x1} is identical in all cases; the same coincidence occurs in \cref{eq:4Q_3barx3}-\cref{eq:4Q_V1x1} and for meson-baryon systems below. We do not have a simple explanation for this coincidence.

The sum of squares of color factors for attractive potentials is largest for the $\mathbf{1}\otimes\mathbf{1}$ color state.
The naive sum of Coulomb energies therefore suggests that it is unlikely that any other color state above leads to a lower-energy state than  $\mathbf{1}\otimes\mathbf{1}$ and therefore that it is unlikely that equal-mass bound hexaquarks arise in pNRQCD at LO or NLO.

This supposition can then be directly tested with GFMC calculations starting from variationally optimized versions of the trial wavefunctions introduced above.
For LO calculations, the only dimensionful parameters are $r$ and $m_Q$, and it is convenient to use ``atomic units'' that have been rescaled by the prefactor of the $1/r$ potential.
In particular, calculations are performed with $\alpha_s = 3/2$ such that the LO prefactor of $1/r$ is equal to unity. 
Results for arbitrary values of $m_Q$ are then obtained by computing $\alpha_s$ using the renormalization scale choice defined by \cref{eq:mup} and then rescaling momenta (energies) by $(2/3) \alpha_s$ (squared) with lengths and imaginary times rescaled by the inverses of these quantities.
At NLO, the renormalization scale enters explicitly, and calculations must be performed independently for each value of $m_Q$.
Apart from the scalings arising at LO, only weak dependence on $m_Q$ is observed in the results, as expected from the logarithmic sensitivity of the NLO potential to $\mu / m_Q$.
For concreteness, results below use a pNRQCD pole mass of $m_b = 4.86831\,\mathrm{GeV}$ corresponding to $\alpha_s(\mu)=0.227325$ based on tuning in Ref.~\cite{Assi:2023cfo}.

For each color state described above, Coulombic trial wavefunctions of the form \cref{eq:PsiH_general} are employed.
The values $a_{ij} = 2/(\mathcal{C}_{ij} \alpha_V)$ are found to be local minima of $\Delta E_{6Q}$, just as in the four-quark studies of Ref.~\cite{Assi:2023dlu}.
For the $\mathbf{1}\otimes \mathbf{1}$ wavefunction, we additionally include Hylleraas-type wavefunctions of the form \cref{eq:PsiH} with the parameter $b/a$ setting the ratio of the ``Bohr radii'' for exponentials involving inter-baryon and intra-baryon quark pairs.
We find that the minimum value of $\Delta E_{6Q}$ is obtained as $b/a \rightarrow \infty$ at both LO and NLO, corresponding to an uncorrelated product of baryons.
In both cases, the energy is equal to twice the single-baryon energies $\Delta E_{QQQ}$.

\begin{figure}[t]
  \centering
  \includegraphics[width=\linewidth]{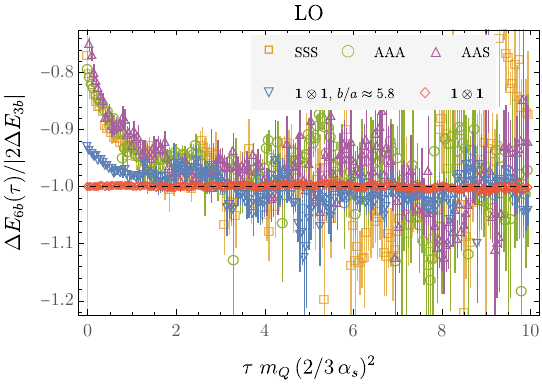}\\
  \includegraphics[width=\linewidth]{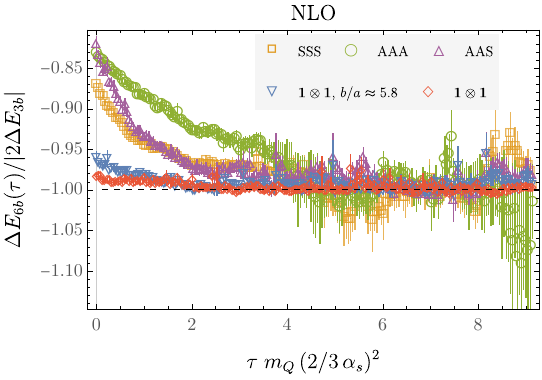}
  \caption{%
    Effective energies for equal‑mass hexaquark trial states with imaginary-time evolution computed using GFMC in ``atomic units'' where the prefactor $(2/3)\alpha_s$ of the color-antisymmetric quark-quark potential is set to unity.  
  \emph{Top:} leading‑order (LO) evolution of the compact $\mathbf{1\otimes1}$ color‑singlet state, the elongated singlet ($b/a\gg1$), and the ${AAA}$, ${AAS}$, and ${SSS}$ color structures defined in the text.  
  \emph{Bottom:} next‑to‑leading‑order (NLO) evolution using $\alpha_s(\mu=m_b)=0.227325$ at $m_b=4.86831\;\mathrm{GeV}$.  
  In both panels, the dashed horizontal line marks the open two‑baryon threshold.%
  }
  \label{fig:hexa_equalmass_effmass}
\end{figure}

Optimized wavefunctions with $AAA$, $AAS$, $SSS$, and $\mathbf{1}\otimes\mathbf{1}$ color structures\footnote{Trial wavefunctions with $ASA$ and $SAA$ color structures give identical results to $AAS$ for equal-mass quarks by permutation symmetry.}, as well as Hylleraas-type wavefunctions with $b/a\approx 5.8$ and $\mathbf{1}\otimes\mathbf{1}$ color-structure describing QED molecules, are used as trial wavefunctions for GFMC calculations.\footnote{The scale $r$ at which $\alpha_V(\mu_p, r)$ is computed to determine the Bohr radii $2/(\mathcal{C}_{ij} \alpha_V)$ is chosen here and below such that $\ln(\mu_p r) = 0.5$ based on single-baryon trial-wavefunction optimization in Ref.~\cite{Assi:2023cfo}.}
Based on results from Ref.~\cite{Assi:2023cfo} for baryons and additional numerical tests for six-quark systems, at LO an imaginary timestep of $\delta \tau = 0.02 / m_Q$ in atomic units is found to lead to negligible discretization effects.
At NLO we take $\delta \tau = 0.5 / m_Q$. 
To ensure that excited-state effects are suppressed, imaginary times $\tau m_Q \gtrsim 10 / (2/3\, \alpha_s )^2$ are employed that are larger than the inverse binding energies arising in the QED analog of positronium molecules.
In particular, we set $N_{\rm steps} = 500$ at LO and $N_{\rm steps} = 200$ at NLO, with $N_{\rm walkers} = 1000$ in both cases.
If there were a bound state with binding energy $\gtrsim 0.1$ in atomic units present, then $O(e^{-\tau (\Delta E_1- \Delta E_0)})$ excited-state suppression would ensure that effective energies from trial wavefunctions with modest ground-state overlaps would approach the bound-state energy in our calculations.
No such bound state is observed.
As seen in \cref{fig:hexa_equalmass_effmass}, effective energies computed using each trial wavefunction converge to $\Delta E_{6Q}(\tau) \approx 2 \Delta E_{QQQ}$ for $\tau m_Q \gtrsim 3$-$5$ in atomic units for $AAA$, $AAS$, and $SSS$ wavefunctions at both LO and NLO.
Trial wavefunctions with $\mathbf{1}\otimes\mathbf{1}$ wavefunctions converge even faster, with $b/a \approx \infty$ wavefunctions apparently converging immediately at LO.
All of these numerical results are consistent with the scenario that $\Delta E_{6Q} = 2 \Delta E_{QQQ}$ and therefore two-baryon systems are unbound at LO and NLO.

\begin{figure}[t]
  \centering
  \includegraphics[width=\linewidth]{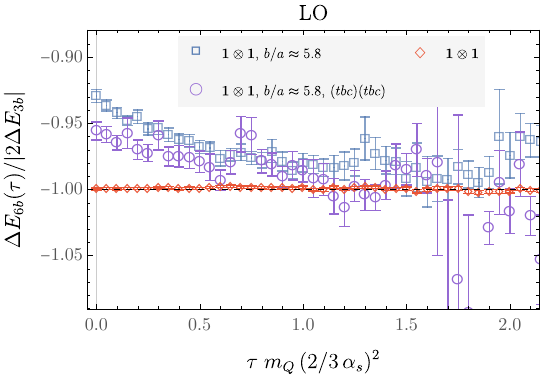}\\
  \caption{%
    LO results for a subset of the trial states in \cref{fig:hexa_equalmass_effmass} compared to those including explicit quark antisymmetry, suitable for a product of two identical spin-1/2 (heavy mass) $(uds)$-flavored baryons.
  }
  \label{fig:hexa_equalmass_effmass_perm}
\end{figure}

The spin-independence of the pNRQCD Hamiltonian at LO and NLO implies that it is not necessary to explicitly incorporate quark antisymmetry into our spatial and color trial wavefunctions.
Simple spin configurations in which one baryon has three spin-up quarks and the other has three spin-down quarks lead to valid total wavefunctions in conjunction with $\mathbf{1}\otimes \mathbf{1}$ trial wavefunctions; for other operators, each diquark should be chosen to be spin-singlet (spin-triplet) if it is color-symmetric (color-antisymmetric).
However, other spin configurations require different trial wavefunctions with explicit antisymmetrization in order to construct valid total wavefunctions.
Such configurations are described by linear combinations of the operators above with color indices and spatial coordinates permuted differently for different terms.
To test the possibility that such a trial wavefunction could lead to a lower-energy state that has nearly zero overlap with the ones studied above, we explicitly constructed linear combinations corresponding to products of spin-1/2 baryons with the correct quark antisymmetrization for a variety of flavor structures such as $(tbb)(ttb)$, $(tbb)(tbb)$, and $(tbc)(tbc)$.
As seen in \cref{fig:hexa_equalmass_effmass_perm} for the $(tbc)(tbc)$ case, only minor differences arise between GFMC results for these wavefunctions and the ones described above.

\subsection{Pentaquarks / meson-baryon systems}

Five-quark $\overline{Q}QQQQ$ states describing bound pentaquarks or unbound meson-baryon systems can be analyzed analogously in pNRQCD.
Interpolating operators for these states can be defined as
\begin{equation}\label{eq:5Q_O_def}
    \mathcal{O}^{(\overline{Q}QQQQ,\mathcal{C})} = \chi_i \psi_j \psi_k \psi_l \psi_m \mathcal{T}^{(\overline{Q}QQQQ,\mathcal{C})}_{ijklm},
\end{equation}
where a complete basis of color tensors is provided by three structures $\mathcal{C} \in \{ \mathbf{1}\otimes\mathbf{1}, \ \mathbf{8}\otimes\mathbf{8}, \  \mathbf{8}_b \otimes\mathbf{8}_b \}$~\cite{Rashmi:2024ako} defined as
\begin{equation}
  \begin{aligned}\label{eq:5Q_1x1_def}
    \mathcal{T}^{(\overline{Q}QQQQ,\mathbf{1}\otimes\mathbf{1})} &= \frac{1}{3\sqrt{2}} \delta_{ij} \varepsilon_{klm}, \\
    \mathcal{T}^{(\overline{Q}QQQQ,\mathbf{8}\otimes\mathbf{8})} &= \frac{1}{2} T^a_{ij} T^a_{nk} \varepsilon_{nlm}, \\
    \mathcal{T}^{(\overline{Q}QQQQ,\mathbf{8}_b\otimes\mathbf{8}_b)} &= \frac{1}{2} T^a_{ik} T^a_{nm} \varepsilon_{njl} .
\end{aligned}
\end{equation}
The potentials for coordinate basis states $\ket{\overline{Q}QQQQ} \equiv \ket{\chi(\bm{x}_1) \psi(\bm{x}_2) \psi(\bm{x}_3) \psi(\bm{x}_3) \psi(\bm{x}_4) \psi(\bm{x}_5)}$ can be defined exactly as in \cref{eq:potential_coords},
\begin{equation}
  V_{\mathcal{C}}^{(\overline{Q}QQQQ)} \equiv \mbraket{ \overline{Q}QQQQ  }{ V^{\psi\chi} + V^{\psi\psi} }{  \overline{Q}QQQQ}.
\end{equation}
For the $\mathbf{1}\otimes\mathbf{1}$ color configuration, the potential at LO and NLO exhibits cancellation of van der Waals potentials between the meson and baryon constituents and takes the simple form,
\begin{equation}\label{eq:5Q_1x1}
\begin{split}
    \frac{V_{\mathbf{1}\otimes\mathbf{1}}^{(\overline{Q}QQQQ,\text{NLO})}}{\alpha_V} &= -\frac{4}{ 3 |\bm{r}_{12}|} \\
    &\hspace{20pt} - \frac{2}{3} \left( \frac{1}{|\bm{r}_{34}|} + \frac{1}{|\bm{r}_{35}|} + \frac{1}{|\bm{r}_{45}|} \right).
    \end{split}
\end{equation}
Again, this occurs from cancellations between attractive and repulsive potentials for each $Q\overline{Q}$ and $QQ$ pair involving one meson and one baryon constituent.

The potentials for the linearly independent color states can be computed similarly.
The $\mathbf{8}\otimes\mathbf{8}$ potential takes the form,
\begin{equation}\label{eq:5Q_8x8}
\begin{split}
    \frac{V_{\mathbf{8}\otimes\mathbf{8}}^{(\overline{Q}QQQQ,\text{NLO})}}{\alpha_V} &= \frac{1}{ 6|\bm{r}_{12}|}    - \frac{1}{6}\left( \frac{7}{ |\bm{r}_{13}|} + \frac{1}{ |\bm{r}_{14}|} + \frac{1}{ |\bm{r}_{15}|} \right) \\
    &\hspace{20pt} - \frac{7}{12} \left( \frac{1}{|\bm{r}_{24}|} + \frac{1}{|\bm{r}_{25}|} \right) \\    
    &\hspace{20pt} + \frac{1}{12} \left( \frac{1}{|\bm{r}_{34}|} + \frac{1}{|\bm{r}_{35}|} \right) \\
    &\hspace{20pt} -\frac{1}{3|\bm{r}_{23}|}  - \frac{2}{3|\bm{r}_{45}|}.
    \end{split}
\end{equation}
The $\mathbf{8}_b\otimes\mathbf{8}_b$ potential is similarly given by
\begin{equation}\label{eq:5Q_8bx8b}
\begin{split}
    \frac{V_{\mathbf{8}\otimes\mathbf{8},b}^{(\overline{Q}QQQQ,\text{NLO})}}{\alpha_V} &= \frac{1}{ 6 |\bm{r}_{13}|} -\frac{1}{6}\left( \frac{7}{ |\bm{r}_{15}|} + \frac{1}{ |\bm{r}_{12}|} + \frac{1}{ |\bm{r}_{14}|}  \right)  \\
    &\hspace{20pt} - \frac{7}{12} \left( \frac{1}{|\bm{r}_{23}|} + \frac{1}{|\bm{r}_{34}|} \right) \\    
    &\hspace{20pt} + \frac{1}{12} \left( \frac{1}{|\bm{r}_{25}|} + \frac{1}{|\bm{r}_{45}|} \right) \\
    &\hspace{20pt} -\frac{1}{3|\bm{r}_{35}|}  - \frac{2}{3|\bm{r}_{24}|}.
    \end{split}
\end{equation}
The $\mathbf{8}_b\otimes\mathbf{8}_b$ potential can be recognized as identical to the $\mathbf{8}\otimes \mathbf{8}$ potential with the quark indices permuted as $(2,3,4,5) \rightarrow (3,5,2,4)$. For equal-mass $\overline{Q}QQQQ$ systems, the $\mathbf{8}_b\otimes\mathbf{8}_b$ potential will therefore provide identical matrix elements to the $\mathbf{8}\otimes \mathbf{8}$ potential.

The sum of the squares of the color factors for the attractive potentials in \cref{eq:5Q_8x8} (and \cref{eq:5Q_8bx8b}) is smaller than the corresponding sum of squares for the color factors in \cref{eq:5Q_1x1}. Within the naive sum of Coulomb energy approximation, this suggests that it is unlikely that the $\mathbf{8}\otimes\mathbf{8}$ (or $\mathbf{8}_b\otimes\mathbf{8}_b$) color configurations will lead to an equal-mass pentaquark bound state.

\begin{figure}[t]
  \includegraphics[width=\linewidth]{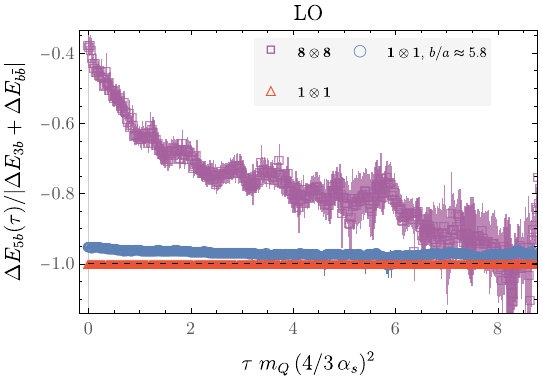} \\
  \includegraphics[width=\linewidth]{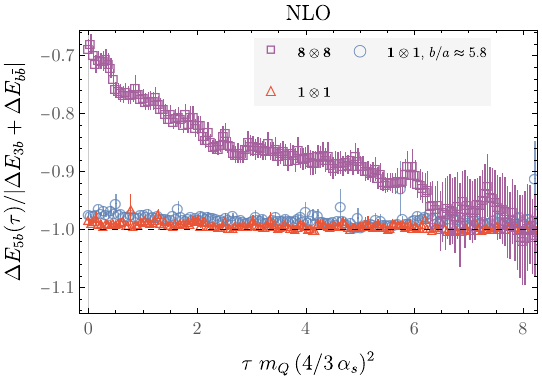} 
  \caption{%
  Top: effective energies for equal‑mass pentaquarks at LO in pNRQCD.  
  Colors and markers distinguish the three trial wavefunctions discussed in the text: the compact color‑singlet configuration $\mathbf{1\otimes 1}$ (red circles), an elongated $\mathbf{1\otimes 1}$ state with aspect ratio $b/a\gg1$ (orange triangles), and the color‑octet configuration $\mathbf{8\otimes 8}$ (green squares).  
  Bottom: analogous results at NLO, obtained with $m_Q = m_b$ and the renormalisation scale from Ref.~\cite{Assi:2023cfo} and $\mu_p$ defined in Eq.~\eqref{eq:mup}.
  In both panels, the dashed horizontal line marks the open meson‑baryon threshold.%
  \label{fig:penta_equalmass_effmass}}
\end{figure}

As in the hexaquark case, we can analyze the pentaquark spectrum more quantitatively by first employing VMC to determine the optimal trial states and then evolving the resulting wavefunctions in imaginary time with GFMC.
Here, LO results use ``atomic units'' for the color-singlet meson-meson potential with $\alpha_s = 3/4$, a step size of $\delta\tau=0.02$, and a propagation length of $N_{\rm steps}=500$.
NLO results use $m_Q = m_b = 4.86831$ GeV as in Ref.~\cite{Assi:2023cfo} with $\delta\tau=0.2$ and $N_{\rm steps}=1000$.
Statistics of $N_{\rm walkers}=1000$ are used in both cases.
In both cases, this is sufficient to ensure that $\tau m_Q \gtrsim 10 / (4/3\, \alpha_s )^2$.
The minimum-energy configuration for $\mathbf{1}\otimes\mathbf{1}$ wavefunctions is found to be an uncorrelated meson-baryon product wavefunction; results for this and a Hylleraas-type wavefunction with $b/a \approx 5.8$ are shown in \cref{fig:penta_equalmass_effmass}.
Effective energies computed with all trial wavefunctions approach the open meson-baryon threshold, although the $\mathbf{8}\otimes \mathbf{8}$ trial states are much noisier and show large fluctuations for large imaginary times.
The GFMC results are consistent with there being no equal-mass meson-baryon bound states.

\section{NNLO van der Waals potentials}\label{sec:NNLO}

\begin{figure*}[tb]
    \centering
    \includegraphics[width=0.32\linewidth]{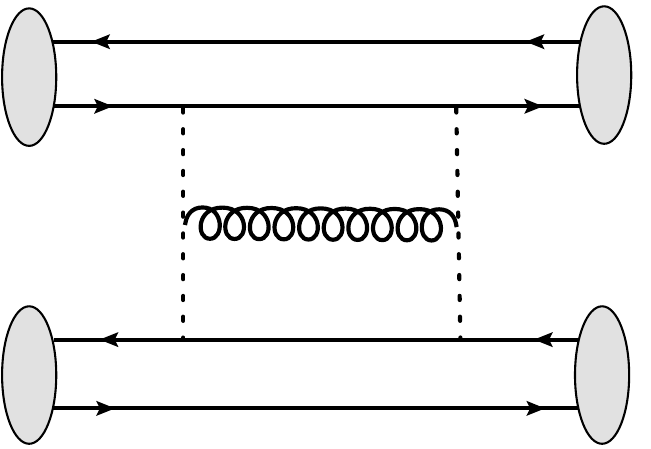}
    \includegraphics[width=0.32\linewidth]{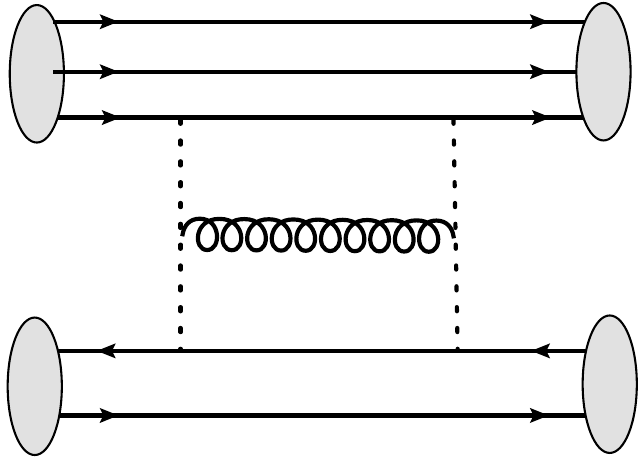}
    \includegraphics[width=0.32\linewidth]{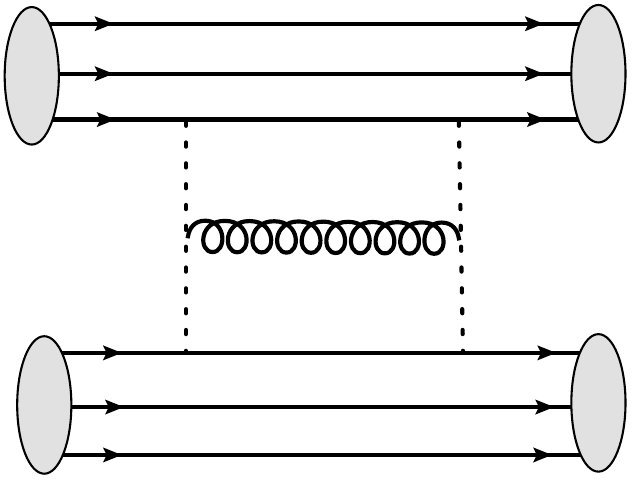}
    \caption{Feynman diagrams with two active quarks that lead to non-vanishing interactions between color-singlet meson-meson (left), meson-baryon (center), and baryon-baryon systems.}
    \label{fig:2body_H}
\end{figure*}

All NNLO contributions to the pNRQCD potential that lead to the same color structure as the LO potential will lead to vanishing van der Waals potentials between color-singlet hadrons due to the same cancellations discussed for LO and NLO above.
In particular, this shows that $1/m_Q$ and $1/m_Q^2$ potentials arising at NNLO do not lead to van der Waals potentials between color-singlet hadrons---any non-vanishing NNLO van der Waals potentials must arise from the static potential and lead to spin- (and flavor-)independent interactions between hadrons.

Such NNLO van der Waals potentials can indeed arise
from two-gluon-exchange diagrams involving gluon self-interactions that have different color structures than the LO potential.
Two-quark, three-quark, and four-quark potentials all make contributions and are considered in turn below.
It is demonstrated that pNRQCD van der Waals potentials are non-zero for some color-singlet hadron-hadron product states.
However, the leading $1/R$ term in a multipole expansion of this van der Waals potential is explicitly demonstrated to vanish, in agreement with the argument from Gauss's law in \cref{sec:Gauss}.

\subsection{Two-quark van der Waals potentials}

In Coulomb gauge, the only diagram that leads to $Q\overline{Q}$ potentials with a different color structure than their LO counterparts is the H-diagram~\cite{Kniehl:2004rk,Collet:2011kq} shown in Fig.~\ref{fig:2body_H}.
The contribution to the potential is $i$ times the value of this diagram, which is given by~\cite{Kummer:1996jz}
\begin{equation}\label{eq:MQQbar}
    \begin{split}
      V^{(Q\overline{Q})}(\bm{r}_{12}) &= \frac{\alpha_s^3}{(4\pi)^2} \mathcal{C}^{(Q\overline{Q})} \mathcal{H}(\bm{r}_{12}), 
    \end{split}
\end{equation}
where the color tensor is
  \begin{equation}
    \mathcal{C}^{(Q\overline{Q})}_{ijkl} = (T^a T^c)_{ij} ([T^b]^T [T^d]^T)_{kl} f^{abe} f^{dce},
  \end{equation}
and the H-diagram spatial integral is given by
\begin{equation}\label{eq:H}
\begin{split}
   \mathcal{H}(\bm{r}_{12}) &= \frac{(4\pi)^5}{2}  \int \frac{d^3q_1 d^3q_2 d^3q_3}{(2\pi)^{9}}  \\
   &\hspace{20pt} \times \left[ (\bm{q}_2 - \bm{q}_1)_i (\delta_{ij} - \bm{k}_i \bm{k}_j /\bm{k}^2)(\bm{q}_3 - \bm{q}_4)_j  \right] \\
   &\hspace{20pt} \times \frac{e^{i\bm{q}_1\cdot \bm{r}_1} e^{i\bm{q}_2\cdot \bm{r}_2} e^{i\bm{q}_3\cdot \bm{r}_1} e^{i\bm{q}_4 \cdot \bm{r}_2} }{ \bm{q}_1^2 \bm{q}_2^2 \bm{q}_3^2 \bm{q}_4^2  \bm{k}^2} \\
  &= \frac{2\pi^2 (12 - \pi^2)}{ |\bm{r}_{12} |},
  \end{split}
\end{equation}
where $\bm{q}_4 = -\bm{q}_1 - \bm{q}_2 - \bm{q}_3$ and $\bm{k} = \bm{q}_1 + \bm{q}_2 - \bm{q}_3 - \bm{q}_4$; see \cref{app:NNLO} for detailed discussion of the Feynman diagram factors involved.
Contracting the color tensor with interpolating operators proportional to $\delta_{ij} \delta_{kl}$ for color-singlet states and $T^a_{ij} T^a_{kl}$ for color-adjoint states correctly reproduces known results~\cite{Collet:2011kq} for $\delta a_2^{(\mathbf{8})}$ in Eq.~\eqref{eq:a2r8} as demonstrated in \cref{app:NNLO}.

The analogous H-diagram contribution to the $QQ$ potential is given by~\cite{Brambilla:2009cd,Kummer:1996jz,Assi:2023cfo}
\begin{equation}\label{eq:MQQ}
    \begin{split}
        V^{(QQ)}(\bm{r}_{12}) &= \frac{\alpha_s^3}{(4\pi)^2} \mathcal{C}^{(QQ)} \mathcal{H}(\bm{r}_{12}), 
    \end{split}
\end{equation}
where the color tensor is
  \begin{equation}
    \mathcal{C}^{(QQ)}_{ijkl} = (T^a T^c)_{ij} (T^b T^d)_{kl} f^{abe} f^{dce}.
  \end{equation}
Contracting this color tensor with interpolating operators for $\overline{\mathbf{3}}$ and $\mathbf{6}$ diquark states reproduces results for $\delta a_2^{(\overline{\mathbf{3}})}$ and $\delta a_2^{(\mathbf{6})}$ from Ref.~\cite{Assi:2023cfo} as demonstrated in \cref{app:NNLO}.

Meson-meson van der Waals forces arising from the $Q\overline{Q}$ and $QQ$ H-diagrams can be derived by contracting the same color tensors with the operator $\mathcal{O}^{(\overline{Q}Q\overline{Q}Q,\mathbf{1}\otimes\mathbf{1})}$ defined in \cref{eq:4Q_O_def,eq:4Q_1x1_def}.
The contractions with both color tensors are identical,
\begin{equation}
  \begin{aligned}
    \mathcal{O}^{(\overline{Q}Q\overline{Q}Q,\mathbf{1}\otimes\mathbf{1})}_{mink} \mathcal{C}^{(QQ)}_{ijkl}  \mathcal{O}^{(\overline{Q}Q\overline{Q}Q,\mathbf{1}\otimes\mathbf{1})}_{mjnl} &= -\frac{2}{3}, \\
    \mathcal{O}^{(\overline{Q}Q\overline{Q}Q,\mathbf{1}\otimes\mathbf{1})}_{mikn} \mathcal{C}^{(Q\overline{Q})}_{ijkl}  \mathcal{O}^{(\overline{Q}Q\overline{Q}Q,\mathbf{1}\otimes\mathbf{1})}_{mjln} &= -\frac{2}{3},
  \end{aligned}
\end{equation}
Meson-baryon and baryon-baryon van der Waals potentials can be computed analogously using the operator defined in \cref{eq:5Q_O_def,eq:5Q_1x1_def} and \cref{eq:6Q_O_def,eq:6Q_1x1_def}, respectively,
\begin{equation}
  \begin{aligned}
    \mathcal{O}^{(\overline{Q}QQQQ,\mathbf{1}\otimes\mathbf{1})}_{minko} \mathcal{C}^{(QQ)}_{ijkl}  \mathcal{O}^{(\overline{Q}QQQQ,\mathbf{1}\otimes\mathbf{1})}_{mjnlo} &= -\frac{2}{3}, \\
    \mathcal{O}^{(\overline{Q}QQQQ,\mathbf{1}\otimes\mathbf{1})}_{kmino} \mathcal{C}^{(Q\overline{Q})}_{ijkl}  \mathcal{O}^{(\overline{Q}QQQQ,\mathbf{1}\otimes\mathbf{1})}_{lmjno} &= -\frac{2}{3}, \\
     \mathcal{O}^{(6Q,\mathbf{1}\otimes\mathbf{1})}_{imnkop} \mathcal{C}^{(QQ)}_{ijkl}  \mathcal{O}^{(6Q,\mathbf{1}\otimes\mathbf{1})}_{jmnlop} &= -\frac{2}{3}.
  \end{aligned}
\end{equation}
In all cases, the hadron-hadron van der Waals potential contribution from two-quark diagrams is given by
\begin{equation}
  V_{HH'}^{(\text{NNLO},2Q)} = -\frac{2}{3} \frac{\alpha_s^3}{(4\pi)^2} \sum_{i \in H, j\in H'} \frac{2\pi^2 (12 - \pi^2)}{ |\bm{r}_{ij} |},
\end{equation}
where $H$ and $H'$ denote the sets of quarks in each hadron.

Consider the limit of widely separated hadrons, i.e. for meson-meson systems $\bm{r}_1, \bm{r}_2 \rightarrow \bm{r}_A$ and $\bm{r}_3, \bm{r}_4 \rightarrow \bm{r}_B$ with 
$|\bm{r}_{12}|, |\bm{r}_{34}| \ll |\bm{r}_{AB}|$. 
Denoting the hadron-hadron separation in this limit by $\bm{R} \equiv \bm{r}_{AB}$, the contributions from quark-quark and quark-antiquark diagrams sum to
\begin{equation}
  V_{MM}^{(\text{NNLO},2Q)} = -\frac{8}{3} \frac{\alpha_s^3}{(4\pi)^2} \frac{2\pi^2 (12 - \pi^2)}{ |\bm{R}|} + O(1/R^2).
\end{equation}
In the limit of large meson-baryon separation where $\bm{r}_1,\bm{r}_2 \rightarrow \bm{r}_A$ and $\bm{r}_3,\bm{r}_4,\bm{r}_5 \rightarrow \bm{r}_B$ the total meson-baryon van der Waals potential contribution is
\begin{equation}
  V_{MB}^{(\text{NNLO},2Q)} = -4 \frac{\alpha_s^3}{(4\pi)^2} \frac{2\pi^2 (12 - \pi^2)}{ |\bm{R} |} + O(1/R^2).
\end{equation}
In the limit of large baryon-baryon separation where $\bm{r}_1,\bm{r}_2,\bm{r}_3 \rightarrow \bm{r}_A$ and $\bm{r}_4,\bm{r}_5,\bm{r}_6 \rightarrow \bm{r}_B$ the total baryon-baryon van der Waals potential contribution is
\begin{equation}
  V_{BB}^{(\text{NNLO},2Q)} = -6 \frac{\alpha_s^3}{(4\pi)^2} \frac{2\pi^2 (12 - \pi^2)}{ |\bm{R} |} + O(1/R^2).
\end{equation}

\subsection{Three-quark van der Waals potentials}

\begin{figure*}[t]
    \centering
    \includegraphics[width=0.67\linewidth]{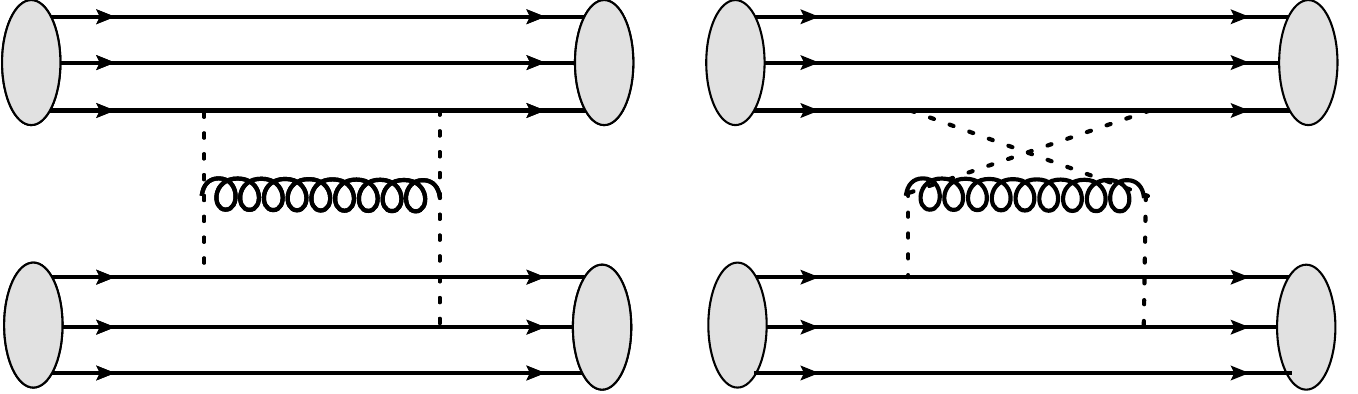}
    \caption{Feynman diagrams with three active quarks that lead to non-vanishing interactions between color-singlet baryon-baryon systems.}
    \label{fig:3body_H}
\end{figure*}

The three-quark potential was derived at NNLO in Ref.~\cite{Brambilla:2009cd}. In Coulomb gauge, it arises from the sum of two diagrams shown in Fig.~\ref{fig:3body_H} and has an amplitude given by
\begin{equation}\label{eq:MQQQ}
    \begin{split}
      V^{(QQQ)}(\bm{r}_{1},\bm{r}_2,\bm{r}_3) &= \frac{\alpha_s^3}{(4\pi)^2} \mathcal{C}^{(QQQ)} v_3(\bm{r}_{12},\bm{r}_{13}), 
    \end{split}
\end{equation}
where the color factor is
\begin{equation}
  \mathcal{C}^{(3Q)}_{ijklmn} \equiv \frac{1}{2} \{T^a, T^c\}_{ij}  T^b_{kl} T^d_{mn} f^{abe} f^{dce},
\end{equation}
and the spatial integral is
  \begin{equation}\label{eq:v3_int}
\begin{split}
  v_3(\bm{r}_{12},\bm{r}_{13}) &= (4\pi)^5  \int \frac{d^3q_1 d^3q_2 d^3q_3}{(2\pi)^{9}}  \\
  &\hspace{5pt} \times \left[ (\bm{q}_2 - \bm{q}_1)_i (\delta_{ij} - \bm{k}_i \bm{k}_j /\bm{k}^2)(\bm{q}_3 - \bm{q}_4)_j  \right] \\
  &\hspace{5pt} \times \frac{e^{i\bm{q}_1\cdot \bm{r}_1} e^{i\bm{q}_2\cdot \bm{r}_2} e^{i\bm{q}_3\cdot \bm{r}_1} e^{i\bm{q}_4 \cdot \bm{r}_3} }{ \bm{q}_1^2 \bm{q}_2^2 \bm{q}_3^2 \bm{q}_4^2  \bm{k}^2},
\end{split}
\end{equation}
where again $\bm{q}_4 = -\bm{q}_1 - \bm{q}_2 - \bm{q}_3$ and $\bm{k} = \bm{q}_1 + \bm{q}_2 - \bm{q}_3 - \bm{q}_4$;
see \cref{app:NNLO} for details of the Feynman diagram factors involved.
A convenient integral representation for computing $v_3(\bm{r},\bm{r}')$ numerically is provided in Ref.~\cite{Brambilla:2009cd}.

To compute the color factors for the baryon-baryon van der Waals potential, first suppose that $\bm{r}_1 = \bm{r}_2$ are in the same baryon.
The first term in the anticommutator in this case becomes
  \begin{equation}\label{eq:non_H_vanishing_B}
    \begin{split}
      & \varepsilon_{ijk}\varepsilon_{i'j'k'} T^a_{il}T^c_{li'} T^b_{jj'} f^{abe}  \\
      &= -\varepsilon_{ijk}\varepsilon_{i'j'k'} T^b_{il}T^c_{li'} T^a_{jj'} f^{abe} \\
      &= \varepsilon_{ijk}\varepsilon_{i'j'k'} T^b_{jl}T^c_{li'} T^a_{ij'} f^{abe} \\
      &= -\varepsilon_{ijk}\varepsilon_{i'j'k'} T^b_{jl}T^c_{lj'} T^a_{ii'} f^{abe},
    \end{split}
  \end{equation} 
and will precisely cancel with the color factor for the diagram where $\bm{r}_1 \leftrightarrow \bm{r}_2$ and therefore $a  \leftrightarrow c$.
  This same cancellation applies to the second piece of the anticommutator, as well as to the case when 
 when $\bm{r}_1 = \bm{r}_3$ are in the same baryon.
Therefore, the only non-vanishing contribution comes when quark $a$ is in one baryon with quarks $b$ and $c$ in the other baryon. All of these are equal by symmetry to
\begin{equation}
    \mathcal{O}^{(6Q,\mathbf{1}\otimes\mathbf{1})}_{ipqkmr} \mathcal{C}^{(3Q)}_{ijklmn}  \mathcal{O}^{(6Q,\mathbf{1}\otimes\mathbf{1})}_{jpklnr} = \frac{1}{3}.
\end{equation}

The limit of large baryon-baryon separation therefore only involves contributions from $V^{(QQQ)}(\bm{r}_{1},\bm{r}_2,\bm{r}_3)$ where $\bm{r}_1 \rightarrow \bm{r}_A$ is in one baryon and $\bm{r}_2,\bm{r}_3 \rightarrow \bm{r}_B$ is in the second baryon.
In this case the spatial integral in \cref{eq:v3_int} can be recognized as twice the H-diagram in \cref{eq:H}, and therefore
\begin{equation}\label{eq:v3_H}
  \begin{split}
    v_3(\bm{r}_{AB},\bm{r}_{AB}) &= 2\mathcal{H}(\bm{r}_{AB}) \\
    &= \frac{4\pi^2 (12 - \pi^2)}{ |\bm{r}_{AB} |}.
\end{split}
\end{equation}
The same result is derived by relating $Q\overline{Q}$ potentials with those for baryons with two quarks in the same position in Refs.~\cite{Brambilla:2009cd,Assi:2023cfo}.
The total contribution to the baryon-baryon NNLO van der Waals potential from three-quark diagrams includes 18 identical contributions of this form,\footnote{There are six choices for which quark corresponds to $\bm{r}_1$. The quarks labeled $\bm{r}_2$ and $\bm{r}_3$ are treated indistinguishably in the sum of the two diagrams, and therefore for each of the six choices of $\bm{r}_1$ there are three independent choices of which quarks in the other baryon are included in the diagram as $\bm{r}_2$ and $\bm{r}_3$.}
  and is therefore given by
\begin{equation}
  V_{BB}^{(\text{NNLO},3Q)} = {\ 12} \frac{\alpha_s^3}{(4\pi)^2} \frac{2\pi^2 (12 - \pi^2)}{ |\bm{R} |} + O(1/R^2).
\end{equation}

A useful cross-check is obtained by evaluating the color-tensor contraction with the single-baryon normalized interpolating operator $1/\sqrt{3} \varepsilon_{ijk}$, which gives
  $\frac{1}{3} \varepsilon_{ikm} \mathcal{C}^{(3Q)}_{ijklmn}  \varepsilon_{jln} =  -1/2  $.
The total NNLO three-quark potential for a color-singlet baryon with $\bm{r}_2 = \bm{r}_3$ is equal to $(-1/2) \alpha_s^3/(4\pi)^2 v_3(\bm{r}_{12},\bm{r}_{12}) = -\alpha_s^3(12-\pi^2)/(8|\bm{r}_{12}|)$ plus singular contributions from diagrams where $\bm{r}_2$ or $\bm{r}_3$ is the quark interacting with two gluon lines,
which agrees with Eq. (51) and Eq. (60) of Ref.~\cite{Brambilla:2009cd}.

The three-quark contribution to meson-meson van der Waals potentials involves diagrams with the same spatial structure and different color tensors,
\begin{equation}
  \begin{aligned}
    \mathcal{C}^{(\overline{Q}\overline{Q}Q)}_{ijklmn} &= \frac{1}{2} (\{[T^a]^T, [T^c]^T\})_{ij}  [T^b]^T_{kl} T^d_{mn} f^{abe} f^{dce}, \\
    \mathcal{C}^{(Q\overline{Q}Q)}_{ijklmn} &= \frac{1}{2} (\{T^a, T^c\})_{ij}  [T^b]^T_{kl} T^d_{mn} f^{abe} f^{dce}.
  \end{aligned}
\end{equation}
Manipulations analogous to \cref{eq:non_H_vanishing_B} show that for meson-meson systems where $\bm{r}_1 = \bm{r}_2$
  \begin{equation}\label{eq:non_H_vanishing_M}
    \begin{split}
      & \delta_{ij}\delta_{i'j'} T^a_{il}T^c_{li'} T^b_{jj'} f^{abe}  \\
      &= -\delta_{ij}\delta_{i'j'} T^b_{il}T^c_{li'} T^a_{jj'} f^{abe} \\
      &= -\delta_{ij}\delta_{i'j'} T^b_{jl}T^c_{li'} T^a_{ij'} f^{abe} \\
      &= -\delta_{ij}\delta_{i'j'} T^b_{jl}T^c_{lj'} T^a_{ii'} f^{abe},
    \end{split}
  \end{equation}
and therefore contractions of these color tensors with meson-meson interpolating operators will cancel with those of the diagram where $\bm{r}_1 \leftrightarrow \bm{r}_2$. 
The only non-vanishing color contractions therefore involve diagrams where $\bm{r}_1 \leftarrow \bm{r}_A$ and $\bm{r}_2,\bm{r}_3 \rightarrow \bm{r}_B$ that are identical to H-diagrams in the large meson-meson separation limit, and \cref{eq:v3_H} applies.
The color contractions in this case are all identical,
\begin{equation}
  \begin{aligned}
    \mathcal{O}^{(\overline{Q}Q\overline{Q}Q,\mathbf{1}\otimes\mathbf{1})}_{imkn} \mathcal{C}^{(\overline{Q}\overline{Q}Q)}_{ijklmn}  \mathcal{O}^{(\overline{Q}Q\overline{Q}Q,\mathbf{1}\otimes\mathbf{1})}_{jmln} &= -\frac{2}{3}, \\
    \mathcal{O}^{(\overline{Q}Q\overline{Q}Q,\mathbf{1}\otimes\mathbf{1})}_{mikm} \mathcal{C}^{(Q\overline{Q}Q)}_{ijklmn}  \mathcal{O}^{(\overline{Q}Q\overline{Q}Q,\mathbf{1}\otimes\mathbf{1})}_{mjln} &= -\frac{2}{3}.
  \end{aligned}
\end{equation}
The overall sign of the diagrams differs from \cref{eq:MQQbar} because there are an odd number of antiquark vertices, and therefore the contribution to the meson-meson van der Waals potential is repulsive, just like in the baryon-baryon case.
There is a counting factor of four corresponding to the choice of which quark interacts with two gluon lines, and therefore, the total contribution to meson-meson van der Waals forces in the large $R$ limit is
\begin{equation}
  V_{MM}^{(\text{NNLO},3Q)} = \frac{16}{3} \frac{\alpha_s^3}{(4\pi)^2} \frac{2\pi^2 (12 - \pi^2)}{ |\bm{R} |} + O(1/R^2).
\end{equation}

For meson-baryon systems, the relevant color tensors are $\mathcal{C}^{(QQQ)}_{ijklmn}$, $\mathcal{C}^{(Q\overline{Q}Q)}_{ijklmn}$, and
\begin{equation}
  \mathcal{C}^{(\overline{Q}QQ)}_{ijklmn} = \frac{1}{2} (\{[T^a]^T, [T^c]^T\})_{ij}  T^b_{kl} T^d_{mn} f^{abe} f^{dce}, \\
\end{equation}
Contractions where $\bm{r}_1$ is in the same hadron as either $\bm{r}_2$ or $\bm{r}_3$ lead to vanishing contributions due to the antisymmetry of \cref{eq:non_H_vanishing_B,eq:non_H_vanishing_M}.
The only non-vanishing contractions are identical to one of
\begin{equation}
  \begin{aligned}
   \mathcal{O}^{(\overline{Q}QQQQ,\mathbf{1}\otimes\mathbf{1})}_{kmiop} \mathcal{C}^{(Q\overline{Q}Q)}_{ijklmn}  \mathcal{O}^{(\overline{Q}QQQQ,\mathbf{1}\otimes\mathbf{1})}_{lnjop} &= -\frac{2}{3}, \\
    \mathcal{O}^{(\overline{Q}QQQQ,\mathbf{1}\otimes\mathbf{1})}_{oikmp} \mathcal{C}^{(QQQ)}_{ijklmn}  \mathcal{O}^{(\overline{Q}QQQQ,\mathbf{1}\otimes\mathbf{1})}_{ojlnp} &= \frac{1}{3}, \\
    \mathcal{O}^{(\overline{Q}QQQQ,\mathbf{1}\otimes\mathbf{1})}_{iokmp} \mathcal{C}^{(\overline{Q}QQ)}_{ijklmn}  \mathcal{O}^{(\overline{Q}QQQQ,\mathbf{1}\otimes\mathbf{1})}_{jolnp} &= \frac{1}{3}.
  \end{aligned}
\end{equation}
There are three independent diagrams where the quark interacting with two gluon lines is in the baryon---these involve $\mathcal{C}^{(Q\overline{Q}Q)}_{ijklmn}$ and have the opposite sign of the $QQQ$ diagram.
There are six independent diagrams where the quark interacting with two gluon lines is in the meson---these involve $\mathcal{C}^{(QQQ)}_{ijklmn}$ or $\mathcal{O}^{(\overline{Q}QQQQ,\mathbf{1}\otimes\mathbf{1})}_{iokmp}$ and have the same sign as the $QQQ$ diagram.
The total contribution to the meson-baryon van der Waals potential in the large $|\bm{R}|$ limit is given by 
\begin{equation}
  V_{MB}^{(\text{NNLO},3Q)} = 8 \frac{\alpha_s^3}{(4\pi)^2} \frac{2\pi^2 (12 - \pi^2)}{ |\bm{R} |} + O(1/R^2).
\end{equation}

\subsection{Four-quark van der Waals potentials}

\begin{figure}[t!]
    \centering
    \includegraphics[width=0.67\linewidth]{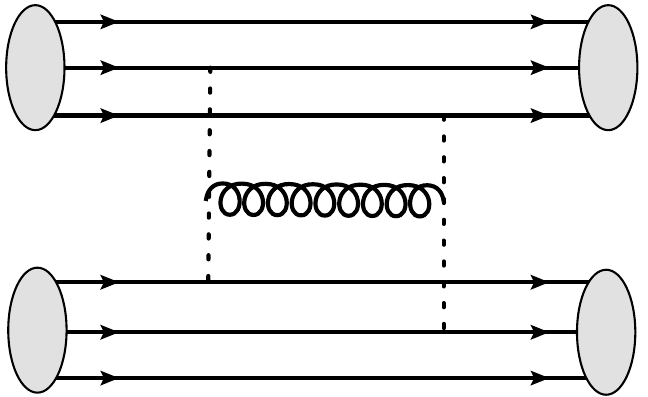}
    \caption{Feynman diagrams with four active quarks that lead to non-vanishing interactions between color-singlet baryon-baryon systems.}
    \label{fig:4body_H}
\end{figure}

It was pointed out in Ref.~\cite{Assi:2023cfo} that four-quark potentials also arise at NNLO for generic multihadron systems. In Coulomb gauge, diagrams involving the four-gluon vertex vanish, and the only diagram that contributes is analogous to the H-diagrams above and shown in \cref{fig:4body_H}.
The contribution to the $QQQQ$ potential from this diagram is
\begin{equation}\label{eq:M4Q}
    \begin{split}
        V^{(4Q)} &= \frac{\alpha_s^3}{(4\pi)^2} \mathcal{C}^{(4Q)} v_4(\bm{r}_1,\bm{r}_2,\bm{r}_3,\bm{r}_4),
    \end{split}
\end{equation}
where the color factor is
  \begin{equation}
    \mathcal{C}^{(4Q)}_{ijklmnop} = T^a_{ij} T^b_{kl} T^c_{mn}  T^d_{op} f^{abe} f^{dce},
  \end{equation}
and the spatial integral is
\begin{equation}\label{eq:v4_int}
\begin{split}
  v_4(\bm{r}_1,\bm{r}_2,\bm{r}_3,\bm{r}_4) &= (4\pi)^5 \int \frac{d^3q_1 d^3q_2 d^3q_3 }{(2\pi)^{9}} \\
  &\hspace{5pt} \times \left[ (\bm{q}_2 - \bm{q}_1)_i (\delta_{ij} - \bm{k}_i \bm{k}_j /\bm{k}^2)(\bm{q}_3 - \bm{q}_4)_j  \right] \\
  &\hspace{5pt} \times \frac{ e^{i\bm{q}_1\cdot \bm{r}_1} e^{i\bm{q}_2\cdot \bm{r}_2} e^{i\bm{q}_3\cdot \bm{r}_3} e^{i\bm{q}_4\cdot \bm{r}_4} }{ \bm{q}_1^2 \bm{q}_2^2 \bm{q}_3^2 \bm{q}_4^2  \bm{k}^2} ,
\end{split}
\end{equation}
where again $\bm{q}_4 = -\bm{q}_1 - \bm{q}_2 - \bm{q}_3$ and $\bm{k} = \bm{q}_1 + \bm{q}_2 - \bm{q}_3 - \bm{q}_4$;
see \cref{app:NNLO} for details of the Feynman diagram factors involved.

If $\bm{r}_1$ and $\bm{r}_2$ correspond to quarks within the same baryon, the contractions of this color-tensor with a color-singlet dibaryon product can be shown to vanish by manipulations analogous to those in \cref{eq:non_H_vanishing_B},
 \begin{equation}\label{eq:H_vanishing_4Q_B}
    \begin{split}
      & \varepsilon_{ijk}\varepsilon_{i'j'k'} T^a_{ii'}T^b_{jj'} f^{abe}  \\
      &= -\varepsilon_{ijk}\varepsilon_{i'j'k'} T^b_{ii'}T^a_{jj'} f^{abe} \\
      &= \varepsilon_{ijk}\varepsilon_{i'j'k'} T^b_{ji'}T^a_{ij'} f^{abe} \\
      &= -\varepsilon_{ijk}\varepsilon_{i'j'k'} T^b_{jj'}T^a_{ii'} f^{abe} \\
      &= 0.
    \end{split}
  \end{equation}
  The same result shows that color-contractions vanish when $\bm{r}_3$ and $\bm{r}_4$ are in the same baryon.
This means that $4Q$ potential contributions to baryon-baryon van der Waals forces only arise from diagrams where $\{\bm{r}_1,\bm{r}_2\}$ correspond to quarks in one baryon and $\{\bm{r}_3,\bm{r}_4\}$ correspond to quarks in the other baryon.
The only non-vanishing contributions in the limit of two widely separated baryons therefore arise from configurations where $\bm{r}_1, \bm{r}_2 \rightarrow \bm{r}_A$ and $\bm{r}_3,\bm{r}_4 \rightarrow \bm{r}_B$.
These are topologically analogous to the two- and three-quark H-diagrams above.
Further, comparing \cref{eq:v4_int} to \cref{eq:H}, the four-quark integral in this limit can be recognized as exactly proportional to the H-diagram spatial integral, 
\begin{equation}
\begin{split}
  v_4(\bm{r}_A,\bm{r}_A,\bm{r}_B,\bm{r}_B) &= v_3(\bm{r}_{AB},\bm{r}_{AB})  = 2\mathcal{H}(\bm{r}_{AB}) \\
    &= \frac{4\pi^2 (12 - \pi^2)}{ |\bm{r}_{AB} |}.
    \end{split}
\end{equation}
Baryon-baryon van der Waals potential contributions in the large $R$ limit can therefore be obtained by multiplying this result with the color-factor contraction
\begin{equation}
    \mathcal{O}^{(6Q,\mathbf{1}\otimes\mathbf{1})}_{ikqmnr} \mathcal{C}^{(4Q)}_{ijklmnop}  \mathcal{O}^{(6Q,\mathbf{1}\otimes\mathbf{1})}_{jlqnpr} = -\frac{1}{6}.
\end{equation}
There are 18 independent contributions of this diagram topology to the baryon-baryon van der Waals potential for a total contribution of
\begin{equation}
  V_{BB}^{(\text{NNLO},4Q)} = -{\ 6} \frac{\alpha_s^3}{(4\pi)^2} \frac{2\pi^2 (12 - \pi^2)}{ |\bm{R} |} + O(1/R^2).
\end{equation}

For meson-meson systems, the topology that does not correspond to an H-diagram involves potential gluon exchange between each color-singlet $\overline{Q}Q$ pair. The color-tensor contraction of this diagram vanishes analogously to \cref{eq:H_vanishing_4Q_B} because
 \begin{equation}\label{eq:H_vanishing_4Q_M}
    \begin{split}
      & \delta_{ij}\delta_{i'j'} T^a_{ii'}T^b_{jj'} f^{abe}  \\
      &= -\delta_{ij}\delta_{i'j'} T^b_{ii'}T^a_{jj'} f^{abe} \\
      &= -\delta_{ij}\delta_{i'j'} T^b_{ji'}T^a_{ij'} f^{abe} \\
      &= -\delta_{ij}\delta_{i'j'} T^b_{jj'}T^a_{ii'} f^{abe} \\
      &= 0.
    \end{split}
  \end{equation}
There are two different H-diagram topologies where potential gluons are exchanged either between $\overline{Q}Q$ pairs in different mesons or between $\overline{Q}\overline{Q}/QQ$ pairs in different mesons,
\begin{equation}
\begin{aligned}
    \mathcal{C}^{(\overline{Q}Q\overline{Q}Q)}_{ijklmnop} &= ([T^a]^T_{ij} T^d_{kl} [T^c]^T_{mn}  T^b_{op}) f^{abe} f^{dce}, \\
    \mathcal{C}^{(\overline{Q}\overline{Q}QQ)}_{ijklmnop} &= ([T^a]^T_{ij} T^c_{kl} [T^b]^T_{mn}  T^d_{op}) f^{abe} f^{dce}.
    \end{aligned}
  \end{equation}
They lead to identical contractions,
\begin{equation}
  \begin{aligned}
    \mathcal{O}^{(\overline{Q}Q\overline{Q}Q,\mathbf{1}\otimes\mathbf{1})}_{iomk} \mathcal{C}^{(\overline{Q}Q\overline{Q}Q)}_{ijklmnop}  \mathcal{O}^{(\overline{Q}Q\overline{Q}Q,\mathbf{1}\otimes\mathbf{1})}_{jpnl} &= -\frac{2}{3}, \\
    \mathcal{O}^{(\overline{Q}Q\overline{Q}Q,\mathbf{1}\otimes\mathbf{1})}_{imko} \mathcal{C}^{(\overline{Q}\overline{Q}QQ)}_{ijklmnop}  \mathcal{O}^{(\overline{Q}Q\overline{Q}Q,\mathbf{1}\otimes\mathbf{1})}_{jnlp} &= -\frac{2}{3}.
  \end{aligned}
\end{equation}
There is only one diagram with each of these color factors, leading to a total contribution to the long-range meson-meson van der Waals potential of
\begin{equation}
  V_{MM}^{(\text{NNLO},4Q)} = -\frac{8}{3} \frac{\alpha_s^3}{(4\pi)^2} \frac{2\pi^2 (12 - \pi^2)}{ |\bm{R} |} + O(1/R^2).
\end{equation}

For meson-baryon systems, diagrams with potential gluons exchanged between the constituents of each hadron again vanish by \cref{eq:H_vanishing_4Q_B,eq:H_vanishing_4Q_M}.
There is one non-vanishing H-diagram topology involves the color factor
\begin{equation}
    \mathcal{C}^{(\overline{Q}QQQ)}_{ijklmnop} \equiv ([T^a]^T_{ij} T^c_{kl} T^b_{mn}  T^d_{op}) f^{abe} f^{dce},
  \end{equation}
which has contraction
\begin{equation}
  \begin{aligned}
    \mathcal{O}^{(\overline{Q}QQQQ,\mathbf{1}\otimes\mathbf{1})}_{iokmq} \mathcal{C}^{(\overline{Q}QQQ)}_{ijklmnop}  \mathcal{O}^{(\overline{Q}QQQQ,\mathbf{1}\otimes\mathbf{1})}_{jplnq} &= \frac{1}{3}.
  \end{aligned}
\end{equation}
The sign is opposite to the baryon-baryon and meson-meson cases, and there are three independent diagrams with this topology. The total long-range contribution to the meson-baryon van der Waals potential is therefore
\begin{equation}
  V_{MB}^{(\text{NNLO},4Q)} = -2 \frac{\alpha_s^3}{(4\pi)^2} \frac{2\pi^2 (12 - \pi^2)}{ |\bm{R} |} + O(1/R^2).
\end{equation}

\subsection{Total}

The total pNRQCD van der Waals potential between hadrons $H$ and $H'$ is given at NNLO by
\begin{equation}
    V_{HH'}^{(\rm NNLO)} \equiv V_{HH'}^{(\rm NNLO, 2Q)} + V_{HH'}^{(\rm NNLO, 3Q)} + V_{HH'}^{(\rm NNLO, 4Q)}.
\end{equation}
The baryon-baryon potential is given by
\begin{equation}
  \begin{split}
    &V_{BB}^{(\rm NNLO)} = -\frac{2}{3}\frac{\alpha_s^3}{(4\pi)^2} \sum_{i \in B_1} \sum_{j \in B_2} \frac{2\pi^2 (12 - \pi^2)}{ |\bm{r}_{ij} |}  \\
    &\hspace{10pt} + \frac{1}{3} \frac{\alpha_s^3}{(4\pi)^2} \left[ \sum_{i \in B_1} \sum_{j<k \in B_2} + \sum_{i \in B_2} \sum_{j<k \in B_1} \right] v_3(\bm{r}_{ij},\bm{r}_{jk}) \\
    &\hspace{10pt} - \frac{1}{3} \frac{\alpha_s^3}{(4\pi)^2} \sum_{i<j \in B_1} \sum_{k<l \in B_2}  v_4(\bm{r}_i,\bm{r}_j,\bm{r}_k,\bm{r}_l),
  \end{split}
\end{equation}
where $B_1 = \{1,2,3\}$ and $B_2 = \{4,5,6\}$ and the last line uses $v_4(\bm{r}_k,\bm{r}_l,\bm{r}_i,\bm{r}_j) = v_4(\bm{r}_i,\bm{r}_j,\bm{r}_k,\bm{r}_l)$ to restrict the sums of four-body potentials to a definite baryon ordering.
The meson-meson potential takes a similar form
\begin{equation}
  \begin{split}
    &V_{MM}^{(\rm NNLO)} = -\frac{2}{3}\frac{\alpha_s^3}{(4\pi)^2} \sum_{i \in M_1} \sum_{j \in M_2} \frac{2\pi^2 (12 - \pi^2)}{ |\bm{r}_{ij} |}  \\
    &\hspace{10pt} + \frac{2}{3} \frac{\alpha_s^3}{(4\pi)^2} \left[ \sum_{i \in M_1} \sum_{j<k \in M_2} + \sum_{i \in M_2} \sum_{j<k \in M_1} \right] v_3(\bm{r}_{ij},\bm{r}_{jk}) \\
    &\hspace{10pt} - \frac{4}{3} \frac{\alpha_s^3}{(4\pi)^2}  v_4(\bm{r}_1,\bm{r}_2,\bm{r}_3,\bm{r}_4),
  \end{split}
\end{equation}
where $M_1 = \{1,2\}$ and $M_2 = \{3,4\}$.
The meson-baryon potential is analogously
\begin{equation}
  \begin{split}
    &V_{MB}^{(\rm NNLO)} = -\frac{2}{3}\frac{\alpha_s^3}{(4\pi)^2} \sum_{i \in M} \sum_{j \in B} \frac{2\pi^2 (12 - \pi^2)}{ |\bm{r}_{ij} |}  \\
    &\hspace{10pt} + \frac{1}{3} \frac{\alpha_s^3}{(4\pi)^2} \left[ \sum_{i \in M} \sum_{j<k \in B} + 2\sum_{i \in B} \sum_{j<k \in M} \right] v_3(\bm{r}_{ij},\bm{r}_{jk}) \\
    &\hspace{10pt} - \frac{2}{3} \frac{\alpha_s^3}{(4\pi)^2} \sum_{i<j \in B_2}  v_4(\bm{r}_1,\bm{r}_2,\bm{r}_i,\bm{r}_j).
  \end{split}
\end{equation}
where $M = \{1,2\}$ and $B = \{3,4,5\}$.

Combining two-, three-, and four-quark contributions leads to a perfect cancellation of the $\mathcal{O}(1/R)$ long-range parts of the baryon-baryon van der Waals potential,
\begin{equation}
\begin{split}
    V_{BB}^{(\rm NNLO)} &= (-6 + 12 -6) \frac{\alpha_s^3}{(4\pi)^2} \\
    &\hspace{20pt} \times \frac{2\pi^2 (12 - \pi^2)}{ |\bm{r}_{AB} |}  + O(1/R^2)\\
    &= 0 + O(1/R^2).
    \end{split}
\end{equation}
Similar cancellations arise for meson-meson systems
\begin{equation}
\begin{split}
    V_{MM}^{(\rm NNLO)} &= \left(-\frac{8}{3} + \frac{16}{3}  - \frac{8}{3} \right) \frac{\alpha_s^3}{(4\pi)^2}\\
    &\hspace{20pt} \times \frac{2\pi^2 (12 - \pi^2)}{ |\bm{r}_{AB} |}  + O(1/R^2)\\
    &= 0 + O(1/R^2),
    \end{split}
\end{equation}
and meson-baryon systems,
\begin{equation}
\begin{split}
    V_{MB}^{(\rm NNLO)}  &= \left(-4 + 8  - 4 \right) \frac{\alpha_s^3}{(4\pi)^2} \\
    &\hspace{20pt} \times \frac{2\pi^2 (12 - \pi^2)}{ |\bm{r}_{AB} |}  + O(1/R^2)\\
    &= 0 + O(1/R^2).
    \end{split}
\end{equation}

The next order of the multipole expansion can be studied using a configuration where one quark in one hadron is separated by a distance $\bm{r}_1$ from the others, while the second hadron is separated by $\bm{R} \gg \bm{r}_1$ and has all quarks at one point.
The arguments for vanishing $O(1/R^2)$ van der Waals potentials in \cref{sec:Gauss} imply a vanishing potential here because $O(1/R^3)$ terms must be proportional to both $r_1$ and the size of the second hadron $r_2 = 0$.
This vanishing can be directly observed from the results above.
Since there are only three independent spatial positions, the four-quark potential contributions simplify to $v_4(\bm{r}_A,\bm{r}_A,\bm{r}_B,\bm{r}_C) = v_3(\bm{r}_{AB},\bm{r}_{AC}) $ or $v_4(\bm{r}_A,\bm{r}_A,\bm{r}_B,\bm{r}_B) = 2 \mathcal{H}(\bm{r}_{AB})$.
This enables exact cancellations between two-, three-, and four-quark potentials. For example, in the meson-meson case 
\begin{equation}
  \begin{split}
    &V_{MM}^{(\rm NNLO)} = -\frac{2}{3}\frac{\alpha_s^3}{(4\pi)^2}\left[ 2 \mathcal{H}(\bm{R}) + 2 \mathcal{H}(\bm{R}+\bm{r}_1) \right]  \\
    &\hspace{10pt} + \frac{2}{3} \frac{\alpha_s^3}{(4\pi)^2} \left[ 2v_3(\bm{R},\bm{R}+\bm{r}_1) + 2 \mathcal{H}(\bm{R}) + 2 \mathcal{H}(\bm{R}+\bm{r}_1) \right]  \\
    &\hspace{10pt} - \frac{4}{3} \frac{\alpha_s^3}{(4\pi)^2}  v_3(\bm{R},\bm{R}+\bm{r}_1) + O(r_1 r_2 / R^3) \\
    &= 0  + O(r_1 r_2 / R^3).
  \end{split}
\end{equation}
Similar cancellations occur in the baryon-baryon and meson-baryon cases.
This illustrates how the absence of $1/R$ and $1/R^2$ potential contributions, forbidden by Gauss's law, is explicitly manifested in pNRQCD at NNLO through cancellations between individually non-vanishing two-, three-, and four-quark potential contributions.

Gauss's law does not forbid higher-order contributions to the multipole expansion.
Non-vanishing dipole-dipole contributions can be expected to arise at NNLO for any configuration with two or more independent quark positions within each hadron so that $v_4(\bm{r}_i,\bm{r}_j,\bm{r}_k,\bm{r}_l)$ does not reduce to a three-body potential form such as $v_3(\bm{r}_{ik},\bm{r}_{il})$.
For such configurations, van der Waals potentials are therefore expected to arise at $O(\alpha_s^3 / R^n)$ with $n\geq 3$,\footnote{Cancellations of additional terms beyond $R^{-2}$ in the multipole expansion can be expected by analogy to QED~\cite{Brambilla:2017ffe}.} and if $\alpha_s \ll 1$ they are too weak to form hadron-hadron bound states by the arguments of \cref{sec:nobind}.
For quark masses where $m_Q \sim \Lambda_{\rm QCD}$ where $\alpha_s \sim O(1)$, this argument breaks down, and bound-state formation through van der Waals potentials cannot be excluded.

\section{Discussion}
\label{sec:conclusion}

In this work, we have argued that baryon-baryon, meson-meson, and meson-baryon systems comprised of equal-mass quarks do not form hadron-hadron bound states when all quark masses are asymptotically large in weakly-coupled pNRQCD.
The steps of the argument can be summarized as follows:
\begin{itemize}
  \item van der Waals potentials between constituents of two different color-singlet hadrons vanish at tree level due to the tracelessness of the generators $T^a$. Hadron-hadron product states are therefore exact eigenstates at LO.
  \item Gauss's law forbids $1/R$ and $1/R^2$ color non-singlet van der Waals potentials. Color-singlet Yukawa potentials $\propto e^{-m_Q R}/R$ are possible but irrelevant when all quark masses are large. 
  \item Van der Waals potentials suppressed by $O(\alpha_s^2 / R^3)$ times powers of $\ln(R)$ are too weak to form bound states from constituents with $O(1/(\alpha_s m_Q))$ radii. Therefore, there are no hadron-hadron bound states with color states described by products of color-singlet hadrons.
    \item Hadron-hadron product states are the lowest-energy color configuration when all hadron constituents have equal masses.
\end{itemize}
The first three steps have been shown analytically and are valid for generic $SU(N)$, $SO(N)$, and $Sp(2N)$ gauge groups relevant to dark sector models.
The last step requires numerical calculations for specific gauge groups and cannot be proven at the same level of rigor. 
Our GFMC calculations of pNRQCD provide evidence that it is valid at LO and NLO for $SU(3)$ gauge theories where all quarks have equal masses satisfying $m_Q \gg \Lambda_{\rm QCD}$.

Several important caveats apply to this weakly-coupled pNRQCD result:
\begin{itemize}
    \item When some quarks are light, two-pion-exchange Yukawa potentials $\propto e^{-2m_\pi R}/R$ take an approximate $1/R$ form for $R \ll 1/(2m_{\pi})$.
    \item Potentials taking an approximate $1/R$ form for $R \ll 1/\Lambda_{\rm QCD}$ might also arise nonperturbatively (e.g. from glueball exchange) that vanish at all orders in the strong-coupling expansion.
    \item For sufficiently light quarks, $\alpha_s \sim O(1)$ and formally NNLO van der Waals potentials could be strong enough to produce bound states.   
\end{itemize}
All three of these caveats point to different possible mechanisms for producing hadron-hadron bound states that should be studied in future work.

These results are relevant to real-world QCD in two distinct ways.
First, they apply to physical $t$, $b$, and $c$ quarks in QCD at NLO, although large corrections to nonrelativistic approximations may arise, especially for $c$ quarks. For $t$ quarks, rapid weak decay rates prevent the formation of even approximately stable hadron bound states in the full Standard Model and are outside the scope of this work.
For $b$ quarks, our numerical results suggest that there are no $(bbb)(bbb)$ hexaquark, $(b\overline{b})(b\overline{b})$ tetraquark, or $(b\overline{b})(bbb)$ pentaquark bound states at LO or NLO in pNRQCD.
On the other hand, it is possible that either ultrasoft effects associated with light quark degrees of freedom that first appear at N$^{3}$LO in pNRQCD could lead to two-pion-exchange Yukawa potentials that are strong enough to form hadron-hadron bound states involving $b$ quarks in QCD.
Lattice NRQCD calculations have searched for $(b\overline{b})(b\overline{b})$ bound states without finding any evidence for them~\cite{Hughes:2017xie}.
On the other hand, lattice NRQCD calculations have suggested the existence of a $(bbb)(bbb)$ bound state with nearly one hundred MeV binding energy~\cite{Mathur:2022ovu}.
It is possible that this dibaryon state arises from light-quark effects, such as two-pion exchange.
Further lattice calculations using a variety of sea quark masses will shed light on mechanisms for forming heavy-hadron bound states, as well as challenging systematic uncertainties related to excited-state effects that have been shown to be unexpectedly large in other calculations of two-baryon systems~\cite{Francis:2018qch,Horz:2020zvv,Amarasinghe:2021lqa,Green:2021qol,Detmold:2024iwz}.

A second connection to real-world QCD arises in conjunction with lattice QCD calculations in which the light quark masses $m_q$ are varied between the physical point where $m_q \ll \Lambda_{\rm QCD}$ and the scale of the strange quark mass where $m_q \sim \Lambda_{\rm QCD}$.
The latter intermediate-quark-mass region should be smoothly connected to the region where all quarks have equal masses $m_q \gg \Lambda_{\rm QCD}$ and can be treated nonrelativistically in pNRQCD.
In the heavy quark mass regime where $m_q \gg \Lambda_{\rm QCD}$, Yukawa potentials are irrelevant, van der Waals potentials are too weak to lead to hadron-hadron bound states, and our GFMC results suggest that hadron-hadron product states are lower energy than all other color configurations.
These results suggest that dibaryon and other hadron-hadron bound states are not present for $m_q \gg \Lambda_{\rm QCD}$ at all orders in the perturbative strong-coupling expansion.

It is not possible to conclude from these results that there are no dibaryon bound states in lattice QCD calculations with moderately heavy quark masses $m_q \sim \Lambda_{\rm QCD}$.
Our results imply that such dibaryon bound states can (only) arise either from effects nonanalytic in $1/m_q$ such as Yukawa potentials associated with light-meson exchange, nonanalytic corrections to the pNRQCD potential, or from the breakdown of the strong coupling expansion when $m_q \sim \Lambda_{\rm QCD}$.
For sufficiently light quark masses such as those arising for $u$ and $d$ quarks in nature, some combination of these nonperturbative relativistic effects presumably give rise to dibaryon bound states such as the deuteron.
Further lattice (NR)QCD studies are needed to map out the quark-mass dependence of hadron-hadron interactions between those of nature where $m_q \ll \Lambda_{\rm QCD}$ and the other extreme limit where $m_q \gg \Lambda_{\rm QCD}$ studied here.

The conclusions that can be drawn from the analytical arguments of this work are that hadron-hadron bound states
cannot arise from van der Waals forces between color-singlet hadrons generated by the pNRQCD potential.
Numerical GFMC calculations further suggest that equal-mass heavy-quark hadron-hadron bound states are unlikely to arise from exotic color configurations of tetraquark, pentaquark, or hexaquark systems.
Hadron-hadron bound states in QCD therefore likely originate from mechanisms beyond the weakly-coupled pNRQCD potential. Possible sources include exotic color configurations that become energetically favorable for sufficiently unequal quark masses or from relativistic effects nonanalytic in $1/m_q$, such as light-meson exchange, relevant only when some quarks are light. In addition, even in the absence of light quarks, nonperturbative dynamics (e.g. glueball exchange) could provide another potential binding mechanism. Determining the precise light quark masses and distance scales where such effects could lead to binding remains a challenging problem.

\begin{acknowledgments}

We thank Yang Bai, Evan Berkowitz, Nora Brambilla, Jakob Hoffmann, Aneesh Manohar, Jerry Miller, Robert Perry, Thomas Richardson, Martin Savage and Antonio Vairo for helpful discussions and their insightful comments. 
This manuscript has been authored by Fermi Forward Discovery Group, LLC under Contract No. 89243024CSC000002 with the U.S. Department of Energy, Office of Science, Office of High Energy Physics.
The work of BA is supported by the DOE grant number DE--SC0011784 and NSF grant number OAC--2103889 as well as the Fermi National Accelerator Laboratory (Fermilab). This manuscript has been authored by Fermi Forward Discovery Group, LLC under Contract No. 89243024CSC000002 with the U.S. Department of Energy, Office of Science, Office of High Energy Physics. BA acknowledges support by the DOE grant de-sc0011784 and NSF  OAC-2103889, OAC-2411215, and OAC-2417682. This work was performed in part at the Aspen Center for Physics, with support for BA by a grant from the Simons Foundation (1161654,Troyer).
\end{acknowledgments}

\appendix

\section{NNLO potential Feynman diagrams}\label{app:NNLO}

Working in Minkowski spacetime and Coulomb gauge, 
$\nabla\!\cdot\!\bm{A}=0$,
the potential ($00$) and transverse ($ij$) gluon propagators decouple,
\begin{equation}\label{eq:CoulombProp}
D_{00}(k)=\frac{i}{\bm{k}^{2}},\quad
D_{ij}(k)=\frac{i\,\delta^{T}_{ij}}
                {E^{2}-\bm{k}^{2}+i\epsilon},\quad
\delta^{T}_{ij}=\delta_{ij}-\frac{\bm{k}_{i}\bm{k}_{j}}{\bm{k}^{2}},
\end{equation}
exactly as employed in the derivation of the pNRQCD
Lagrangian\,\cite{Pineda:1997bj,Brambilla:2004jw}. If two legs are potential ($\mu=\nu=0$) and the third is
transverse ($\rho=i\neq0$), the QCD three-gluon vertex reduces to
\begin{equation}\label{eq:3gCoulomb}
\Gamma^{00i}_{abc}(l_{a},l_{b})=-\,g\,f^{abc}\bigl(l^{i}_{b}-l^{i}_{a}\bigr),
\end{equation}
with all momenta taken incoming.

For static sources, $v^{\mu}=(1,\mathbf{0})$ and
substituting $\gamma^{0}\to\pm1$ for heavy quark/antiquark spinors yields
\begin{equation}
V_{Q Q g}=ig\,T^{a},\quad
V_{\bar Q\bar Q g}=-ig\,T^{a},\quad S_{Q}(k)=\frac{i}{k^{0}+i\epsilon}.
\end{equation}
Using color algebra relations
$T^{a}_{ik}T^{a}_{jl}=\tfrac12
(\delta_{il}\delta_{kj}-\tfrac13\delta_{ik}\delta_{jl})$,
\begin{equation}
\delta_{ij}\delta_{kl}\,T^{a}_{ik}T^{a}_{jl}=\frac43,\qquad
\epsilon_{ijm}\epsilon_{klm}\,T^{a}_{ik}T^{a}_{jl}=-\frac23.
\end{equation}

Combining signs from 
$i\mathcal{M}=-iV$,
the propagator in~\eqref{eq:CoulombProp},
and the vertex factors above give a single Coulomb-exchange between two heavy color
sources generating the potentials
\begin{align}\label{eq:potentials}
V_{Q\bar Q}(\bm{k}) &=
  -\frac{4\pi\alpha_s}{\bm{k}^{2}}
   \Bigl(
       \tfrac{4}{3}\,\mathcal{P}^{(1)}
     - \tfrac{1}{6}\,\mathcal{P}^{(8)}
   \Bigr),\\[6pt]
V_{Q Q}(\bm{k}) &=
   \frac{4\pi\alpha_s}{\bm{k}^{2}}
   \Bigl(
       \tfrac{2}{3}\,\mathcal{P}^{(\overline{3})}
     - \tfrac{1}{3}\,\mathcal{P}^{(6)}
   \Bigr).
\end{align}
The first line splits the quark–antiquark potential into its attractive
color-singlet and repulsive color-octet components, while the second
line separates the diquark potential into the attractive
antisymmetric $\overline{\mathbf 3}$ and the repulsive symmetric
$\mathbf 6$ channels.

The color projectors that appear in Eq.\,\eqref{eq:potentials} are
\begin{align}\label{eq:projectors}
\mathcal{P}^{(1)}_{ij,kl}            &= \tfrac13\,\delta_{ij}\,\delta_{kl}, &
\mathcal{P}^{(\overline{3})}_{ij,kl} &= \tfrac12\bigl(\delta_{ik}\,\delta_{jl}-\delta_{il}\,\delta_{jk}\bigr),\\[4pt]
 \mathcal{P}^{(8)}_{ij,kl}            &= 2\,T^{a}_{il}\,T^{a}_{kj},&
\mathcal{P}^{(6)}_{ij,kl} &= \tfrac12\bigl(\delta_{ik}\,\delta_{jl}+\delta_{il}\,\delta_{jk}\bigr)
-\tfrac13\,\delta_{ij}\,\delta_{kl},
\end{align}
which reproduces the familiar attractive
color-singlet Coulomb potential, together with the repulsive octet
piece in the $Q\bar Q$ sector, and the attractive
$\overline{\mathbf 3}$ versus repulsive $\mathbf 6$ potentials in
the $QQ$ sector.

The four-quark H-diagram shown in \cref{fig:4body_H} is a tree diagram and involves zero energy transfer through all propagators.
Labeling the four potential gluon momenta by $\bm{q}_1$, $\bm{q}_2$, $\bm{q}_3$, and $\bm{q}_4 = \bm{q}_1 - \bm{q}_2 - \bm{q}_3$ and the transverse gluon momentum by $\bm{k} = \bm{q}_1 + \bm{q}_2 - \bm{q}_3 - \bm{q}_4$, the contribution to the potential from this diagram is given by
\begin{widetext}
    \begin{equation}
\begin{split}
  V^{(4Q)} &= (-i)i \int \frac{d^3q_1 d^3q_2 d^3q_3}{(2\pi)^{9}} \left[ (ig T^a) \otimes (ig T^b)  \otimes  (ig T^c) \otimes (ig T^d)  \right] (-gf^{abe})  (-gf^{dce}) \\
  &\hspace{20pt} \times \left[ (\bm{q}_2 - \bm{q}_1)^i (\delta_{ij} - \bm{k}_i \bm{k}_j /\bm{k}^2) (\bm{q}_3 - \bm{q}_4)^j  \right] \frac{e^{i\bm{q}_1\cdot \bm{r}_1} e^{i\bm{q}_2\cdot \bm{r}_2} e^{i\bm{q}_3\cdot \bm{r}_3} e^{i\bm{q}_4\cdot \bm{r}_4}   }{ \bm{q}_1^2 \bm{q}_2^2 \bm{q}_3^2 \bm{q}_4^2  \bm{k}^2},
\end{split}
\end{equation}
where the first factor of $-i$ is from the transfer gluon propagator and the $i$ from relating the potential to the amplitude $-iV$. This can be immediately recognized as \eqref{eq:M4Q} after separating the color factors from the spatial integral and expressing the couplings as $g^6 = (4\pi)^5 \alpha_s^3/(4\pi)^2$.

There are two non-vanishing versions of the $3Q$ H-diagram as shown in \cref{fig:3body_H}, a crossed and an uncrossed one.
The uncrossed one is 
    \begin{equation}
\begin{split}
  V^{(3Q,\text{uncrossed})} &= (-i)i \int \frac{d^3q_1 d^3q_2 d^3q_3 dE}{(2\pi)^{10}} \left[ (ig T^a)(ig T^c)  \otimes  (ig T^b) \otimes (ig T^d)  \right] (-gf^{abe})  (-gf^{dce}) \\
  &\hspace{20pt} \times \left[ (\bm{q}_2 - \bm{q}_1)^i (\delta_{ij} - \bm{k}_i \bm{k}_j /\bm{k}^2) (\bm{q}_3 - \bm{q}_4)^j  \right] \frac{e^{i\bm{q}_1\cdot \bm{r}_1} e^{i\bm{q}_2\cdot \bm{r}_2} e^{i\bm{q}_3\cdot \bm{r}_1} e^{i\bm{q}_4\cdot \bm{r}_3}   }{ \bm{q}_1^2 \bm{q}_2^2 \bm{q}_3^2 \bm{q}_4^2  [-E^2 + \bm{k}^2 - i \epsilon]} \left( \frac{i}{E + i\epsilon} \right),
\end{split}
\end{equation}
while $V^{(3Q,\text{crossed})}$ only differs in the order of $T^a$ and $T^c$.
The full three-quark amplitude is then
\begin{equation}
\begin{split}
  V^{(3Q)} &=  V^{(3Q,\text{uncrossed})}  + V^{(3Q,\text{crossed})} \\
  &= (\{T^a, T^c\} \otimes T^b \otimes T^d) f^{abe} f^{dce} g^6 \int \frac{d^3q_1 d^3q_2 d^3q_3 }{(2\pi)^{12}}  \left[ (\bm{q}_2 - \bm{q}_1)^i (\delta_{ij} - \bm{k}_i \bm{k}_j /\bm{k}^2)(\bm{q}_3 - \bm{q}_4)^j  \right] \\
  &\hspace{20pt} \times \frac{e^{i\bm{q}_1\cdot \bm{r}_1} e^{i\bm{q}_2\cdot \bm{r}_2} e^{i\bm{q}_3\cdot \bm{r}_1} e^{i\bm{q}_4 \cdot \bm{r}_3} }{ \bm{q}_1^2 \bm{q}_2^2 \bm{q}_3^2 \bm{q}_4^2 } \int \frac{dE}{2\pi} \frac{1}{[-E^2 + \bm{k}^2 - i \epsilon]} \left( \frac{i}{E + i\epsilon} \right).
  \end{split}
  \end{equation}
Separating the color factor $\mathcal{C}^{(3Q)}_{ijklmn} = \frac{1}{2} \{T^a, T^c\}_{ij}  T^b_{kl} T^d_{mn} f^{abe} f^{dce}$
and evaluating the energy integral gives
\begin{equation}
\begin{split}
  V^{(3Q)} &= \mathcal{C}^{(3Q)} g^6  \int \frac{d^3q_1 d^3q_2 d^3q_3 }{(2\pi)^{9}}  \left[ (\bm{q}_2 - \bm{q}_1)_i (\delta_{ij} - \bm{k}_i \bm{k}_j /\bm{k}^2)(\bm{q}_3 - \bm{q}_4)_j  \right]  \frac{e^{i\bm{q}_1\cdot \bm{r}_1} e^{i\bm{q}_2\cdot \bm{r}_2} e^{i\bm{q}_3\cdot \bm{r}_1} e^{i\bm{q}_4 \cdot \bm{r}_3} }{ \bm{q}_1^2 \bm{q}_2^2 \bm{q}_3^2 \bm{q}_4^2  \bm{k}^2}.
\end{split}
\end{equation}
This can be immediately recognized as \cref{eq:MQQQ}.

The quark-antiquark H-diagram in \cref{fig:2body_H} can be expressed similarly as
    \begin{equation}
\begin{split}
  V^{(Q\overline{Q})} &= (-i)i \int \frac{d^3q_1 d^3q_2 d^3q_3 dE dF}{(2\pi)^{11}} \left[ (ig T^a)(ig T^c)  \otimes  (-ig [T^b]^T)(ig [T^d]^T)  \right] (-gf^{abe})  (-gf^{dce}) \\
  &\hspace{20pt} \times \left[ (\bm{q}_2 - \bm{q}_1)^i (\delta_{ij} - \bm{k}_i \bm{k}_j /\bm{k}^2) (\bm{q}_3 - \bm{q}_4)^j  \right] \frac{e^{i\bm{q}_1\cdot \bm{r}_1} e^{i\bm{q}_2\cdot \bm{r}_2} e^{i\bm{q}_3\cdot \bm{r}_1} e^{i\bm{q}_4\cdot \bm{r}_2}   }{ \bm{q}_1^2 \bm{q}_2^2 \bm{q}_3^2 \bm{q}_4^2  [-E^2 + \bm{k}^2 - i \epsilon]} \left( \frac{i}{E - F + i\epsilon} \right) \left( \frac{i}{F + i\epsilon} \right),
\end{split}
\end{equation}
Introducing the color factor
$\mathcal{C}^{(Q\overline{Q})}_{ijkl} = (T^a T^c)_{ij} ([T^b]^T [T^d]^T)_{kl} f^{abe} f^{dce}$ and evaluating the energy integrals gives
    \begin{equation}
\begin{split}
  V^{(Q\overline{Q})} &= -\mathcal{C}^{(Q\overline{Q})}  g^6 \int \frac{d^3q_1 d^3q_2 d^3q_3 }{(2\pi)^{10}}  \left[ (\bm{q}_2 - \bm{q}_1)^i (\delta_{ij} - \bm{k}_i \bm{k}_j /\bm{k}^2)(\bm{q}_3 - \bm{q}_4)^j  \right] \\
  &\hspace{20pt} \times \frac{e^{i\bm{q}_1\cdot \bm{r}_1} e^{i\bm{q}_2\cdot \bm{r}_2} e^{i\bm{q}_3\cdot \bm{r}_1} e^{i\bm{q}_4 \cdot \bm{r}_2} }{ \bm{q}_1^2 \bm{q}_2^2 \bm{q}_3^2 \bm{q}_4^2 } \int \frac{dE}{2\pi} \frac{dF}{2\pi} \frac{1}{[-E^2 + \bm{k}^2 - i \epsilon]} \left( \frac{1}{E - F + i\epsilon} \right) \left( \frac{1}{F + i\epsilon} \right) \\
   &= -\mathcal{C}^{(Q\overline{Q})}  g^6 \int \frac{d^3q_1 d^3q_2 d^3q_3 }{(2\pi)^{9}}  \left[ (\bm{q}_2 - \bm{q}_1)^i (\delta_{ij} - \bm{k}_i \bm{k}_j /\bm{k}^2)(\bm{q}_3 - \bm{q}_4)^j  \right] \\
  &\hspace{20pt} \times \frac{e^{i\bm{q}_1\cdot \bm{r}_1} e^{i\bm{q}_2\cdot \bm{r}_2} e^{i\bm{q}_3\cdot \bm{r}_1} e^{i\bm{q}_4 \cdot \bm{r}_2} }{ \bm{q}_1^2 \bm{q}_2^2 \bm{q}_3^2 \bm{q}_4^2 } \int \frac{dE}{2\pi} \frac{1}{[-E^2 + \bm{k}^2 - i \epsilon]}  \left( \frac{-i}{E + i\epsilon} \right) \\
  &= \mathcal{C}^{(Q\overline{Q})} g^6  \frac{1}{2}  \int \frac{d^3q_1 d^3q_2 d^3q_3}{(2\pi)^{9}}  \left[ (\bm{q}_2 - \bm{q}_1)^i (\delta_{ij} - \bm{k}_i \bm{k}_j /\bm{k}^2)(\bm{q}_3 - \bm{q}_4)^j  \right] \frac{e^{i\bm{q}_1\cdot \bm{r}_1} e^{i\bm{q}_2\cdot \bm{r}_2} e^{i\bm{q}_3\cdot \bm{r}_1} e^{i\bm{q}_4 \cdot \bm{r}_2} }{ \bm{q}_1^2 \bm{q}_2^2 \bm{q}_3^2 \bm{q}_4^2  \bm{k}^2},
\end{split}
\end{equation}
which can be immediately recognized as \cref{eq:MQQbar}.

These results imply that three- and four-quark van der Waals potentials in the large hadron-hadron separation limit discussed in the main text are proportional to the quark-antiquark H-diagram, which has been studied extensively in the literature~\cite{Kummer:1996jz,Kniehl:2004rk,Collet:2011kq} with the result that $V^{(Q\overline{Q})}(r) = \mathcal{C}^{(Q\overline{Q})} \frac{\alpha_s^3}{(4\pi)^2}\mathcal{H}(r)$ where $\mathcal{H}(r) = 2\pi^2(12-\pi^2)/r$.

\end{widetext}

Matrix elements of $V^{(Q\overline{Q})}$ can be computed straightforwardly for unit-normalized color-singlet and color-octet $Q\overline{Q}$ states by contracting with
\begin{equation}
    \mathcal{O}^{(\bm{1})}_{ij} = \frac{1}{\sqrt{3}} \delta_{ij},
\end{equation}
and
\begin{equation}
    \mathcal{O}^{(\bm{8})}_{ij} = \sqrt{2} T^1_{ij},
\end{equation}
respectively (defining the operator with other component of $\sqrt{2} T^a_{ij}$ would give equivalent results below).
The difference between the ratios of these contractions to those of the tree-level color-factor $T^A_{ij} T^A_{lk}$ defines the representation-dependent NNLO term in the octet potential $\delta a_2^{(\mathbf{8})}$,
\begin{equation}
\begin{split}
    \delta a_2^{(\bm{8})} &\equiv \left( \frac{ \mathcal{O}^{(\bm{8})}_{ik} \mathcal{C}^{(Q\overline{Q})}_{ijkl} \mathcal{O}^{(\bm{8})}_{jl} }{\mathcal{O}^{(\bm{8})}_{ik} T^A_{ij} T^A_{lk} \mathcal{O}^{(\bm{8})}_{jl} } - \frac{ \mathcal{O}^{(\bm{1})}_{ik} \mathcal{C}^{(Q\overline{Q})}_{ijkl} \mathcal{O}^{(\bm{1})}_{jl} }{\mathcal{O}^{(\bm{1})}_{ik} T^A_{ij} T^A_{lk} \mathcal{O}^{(\bm{1})}_{jl} }  \right) \mathcal{H} \\
    &= \left( \frac{ -3/8 }{  1/6 } - \frac{ -3 }{ -4/3 }  \right) \mathcal{H} \\
    &= -9 \pi^2 (12 - \pi^2),
    \end{split}
\end{equation}
which matches the results of Refs.~\cite{Collet:2011kq,Kniehl:2004rk}. 

Analogous cross-checks can be performed for the color-antisymmetric and color-symmetric $QQ$ potentials.
Unit-normalized operators for these channels can be defined by
\begin{equation}
    \mathcal{O}^{(\overline{\mathbf{3}})}_{ij} = \frac{1}{\sqrt{3}} \varepsilon_{ij1},
\end{equation}
and~\cite{Brambilla:2005yk}
\begin{equation}
    \mathcal{O}^{(\mathbf{6})}_{ij} = \frac{1}{\sqrt{6}} \begin{pmatrix} 1 & 1/\sqrt{2} & 1/\sqrt{2} \\  1/\sqrt{2} & 1 & 1/\sqrt{2} \\  1/\sqrt{2} & 1/\sqrt{2} & 1 \end{pmatrix}_{ij},
\end{equation}
respectively.
The representation-dependent piece of the two-loop color-antisymmetric $QQ$ potential is
\begin{equation}
\begin{split}
    \delta a_2^{(\overline{\mathbf{3}})} &\equiv \left( \frac{ \mathcal{O}^{(\overline{\mathbf{3}})}_{ik} \mathcal{C}^{(QQ)}_{ijkl} \mathcal{O}^{(\overline{\mathbf{3}})}_{jl} }{\mathcal{O}^{(\overline{\mathbf{3}})}_{ik} T^A_{ij} T^A_{kl} \mathcal{O}^{(\overline{\mathbf{3}})}_{jl} } - \frac{ \mathcal{O}^{(\bm{1})}_{ik} \mathcal{C}^{(Q\overline{Q})}_{ijkl} \mathcal{O}^{(\bm{1})}_{jl} }{\mathcal{O}^{(\bm{1})}_{ik} T^A_{ij} T^A_{lk} \mathcal{O}^{(\bm{1})}_{jl} }  \right) \mathcal{H} \\
    &= \left( \frac{ -1 }{  -2/3 } - \frac{ -3 }{ -4/3 }  \right) \mathcal{H} \\
    &= -\frac{3}{2} \pi^2 (12 - \pi^2),
    \end{split}
\end{equation}
and the corresponding result for the color-symmetric $QQ$ potential is
\begin{equation}
\begin{split}
    \delta a_2^{(\mathbf{6})} &\equiv \left( \frac{ \mathcal{O}^{(\mathbf{6})}_{ik} \mathcal{C}^{(QQ)}_{ijkl} \mathcal{O}^{(\mathbf{6})}_{jl} }{\mathcal{O}^{(\mathbf{6})}_{ik} T^A_{ij} T^A_{kl} \mathcal{O}^{(\mathbf{6})}_{jl} } - \frac{ \mathcal{O}^{(\bm{1})}_{ik} \mathcal{C}^{(Q\overline{Q})}_{ijkl} \mathcal{O}^{(\bm{1})}_{jl} }{\mathcal{O}^{(\bm{1})}_{ik} T^A_{ij} T^A_{lk} \mathcal{O}^{(\bm{1})}_{jl} }  \right) \mathcal{H} \\
    &= \left( \frac{ -1/2 }{  1/3 } - \frac{ -3 }{ -4/3 }  \right) \mathcal{H} \\
    &= -\frac{15}{2} \pi^2 (12 - \pi^2),
    \end{split}
\end{equation}
both of which match the results in Ref.~\cite{Assi:2023cfo}.

\bibliography{pnrqcd}

\clearpage

\end{document}